\documentclass[letterpaper,journal]{IEEEtran}

\usepackage[cmex10]{amsmath} 
\usepackage{amsfonts,stmaryrd,acronym,amssymb,amsthm,dsfont}
\usepackage{algorithm}
\usepackage{algorithmic}

\usepackage[font=footnotesize, justification=raggedright, singlelinecheck=false]{caption}
\usepackage[font=footnotesize, justification=centering, singlelinecheck=false]{subcaption}

\usepackage{graphicx,paralist,mdframed}
\usepackage{graphics}
\usepackage{bm}
\usepackage{url}
\usepackage{cite}
\usepackage{caption}
\usepackage{subcaption}
\usepackage{hyperref}
\usepackage{flushend}
\usepackage{enumerate}
\usepackage{pdflscape}
\usepackage[dvipsnames]{xcolor}
\usepackage{mathrsfs}
\usepackage{lipsum}
\usepackage{diagbox}
\usepackage{soul}
\usepackage{kantlipsum}
\usepackage{ragged2e} 
\usepackage{changepage}

\usepackage[bbgreekl]{mathbbol}
\DeclareSymbolFontAlphabet{\mathbbm}{bbold}
\DeclareSymbolFontAlphabet{\mathbb}{AMSb}%

\graphicspath{{Graphics/}}
\definecolor{ColorHighlight}{rgb}{1,0,0}

\makeatletter
\def\blfootnote{\xdef\@thefnmark{}\@footnotetext}
\makeatother
\newcommand{\indi}[1]{\ensuremath{\mathds{1}}}

\newcounter{mytempeqcounter}

\allowdisplaybreaks
\allowbreak

\makeatletter%
\if@twocolumn%
\else
\fi%
\makeatother%

\newcommand{\indic}[1]{\ensuremath{\mathds{1}}}

\renewcommand{\leq}{\leqslant} 
\renewcommand{\geq}{\geqslant} 

\acrodef{ACDIS}[ACDIS]{Adaptive Communication Decision and Information Systems}
\acrodef{AEP}{Asymptotic Equipartition Property}
\acrodef{AoA}{Angle of Arrival}
\acrodef{AWGN}{additive white Gaussian noise}
\acrodef{AVC}[AVC]{Arbitrarily Varying Channel}
\acrodefplural{AVC}{Arbitrarily Varying Channels}
\acrodef{PIR-PNSI}{Private Information Retrieval with Private Noisy Side Information}
\acrodef{BER}{Bit-Error-Rate}
\acrodef{BEC}{Binary Erasure Channel}
\acrodefplural{BEC}{Binary Erasure Channels}
\acrodef{BSC}{Binary Symmetric Channel}
\acrodefplural{BSC}{Binary Symmetric Channels}
\acrodef{BPSK}{Binary Phase-Shift Keying}
\acrodef{BICM}[BICM]{Bit-Interleaved Coded-Modulation}
\acrodef{CDF}[CDF]{Cumulative Distribution Function}
\acrodef{CGF}[CGF]{Cumulant Generating Function}
\acrodef{CLT}[CLT]{Central Limit Theorem}
\acrodef{CSI}[CSI]{channel state information}
\acrodef{DMC}[DMC]{Discrete Memoryless Channel}
\acrodefplural{DMC}{Discrete Memoryless Channels}
\acrodef{DMS}[DMS]{Discrete Memoryless Source}
\acrodef{ERM}[ERM]{Empirical Risk Minimization}
\acrodef{FER}[FER]{Frame Error Rate}
\acrodef{ICA}[ICA]{Independent Component Analysis}
\acrodef{iid}[i.i.d.]{independent and identically distributed}
\acrodef{IoT}[IoT]{Internet of Things}
\acrodef{KKT}[KKT]{Karush-Kuhn Tucker}
\acrodef{LASSO}[LASSO]{Least Absolute Shrinkage and Selection Operator}
\acrodef{LPD}[LPD]{Low Probability of Detection}
\acrodef{LDPC}[LDPC]{Low-Density Parity-Check}
\acrodef{CSCG}[CSCG]{Circularly Symmetric Complex Gaussian Distribution}
\acrodef{LLMS}[LLMS]{Linear Least Mean Square}
\acrodef{LMS}[LMS]{Least Mean Square}
\acrodef{MAC}[MAC]{multiple-Access Channel}
\acrodef{ADSI}[ADSI]{Action-Dependent State Information}
\acrodef{MGF}[MGF]{Moment Generating Function}
\acrodef{MLC}[MLC]{Multi-Level Coding}
\acrodef{MLE}[MLE]{Maximum Likelihood Estimate}
\acrodef{MSD}[MSD]{Multi-Stage Decoding}
\acrodef{MMSE}[MMSE]{minimum mean-square error}
\acrodef{PAC}[PAC]{Probably Approximately Correct}
\acrodef{PCA}[PCA]{Principal Component Analysis}
\acrodef{PDF}[PDF]{probability density function}
\acrodefplural{PDF}{probability density functions}
\acrodef{PMF}[PMF]{Probability Mass Function}
\acrodefplural{PMF}{Probability Mass Functions}
\acrodef{PPM}[PPM]{Pulse Position Modulation}
\acrodef{PSD}{Power Spectral Density}
\acrodef{PSK}{Phase Shift Keying}
\acrodef{QKD}{Quantum Key Distribution}
\acrodef{ROC}{Receiver Operating Characteristic}
\acrodef{CVQKD}{Continuous-Variable \ac{QKD}}
\acrodef{QPSK}{Quadrature Phase-Shift Keying}
\acrodef{RV}{random variable}
\acrodef{SIMO}{Single-Input Multiple-Output}
\acrodef{SNR}{signal-to-noise ratio}
\acrodef{SVM}[SVM]{Support Vector Machine}
\acrodef{TPCP}{Trace-Preserving Completely-Positive}
\acrodef{wrt}[w.r.t.]{with respect to}
\acrodef{WSS}{Wide Sense Stationary}
\acrodef{RHS}{Right Hand Side}
\acrodef{LHS}{Left Hand Side}
\acrodef{PIR}{Private Information Retrieval}
\acrodef{MDS}{Maximum Distance Separable}
\acrodef{LLN}{Law of Large Numbers}
\acrodef{DFRC}{Dual-Function Radar Communication}
\acrodef{ISAC}{Integrated Sensing and Communication}
\acrodef{RadCom}{Joint Radar and Communicatins}
\acrodef{PLS}[PLS]{Physical Layer Security}
\acrodef{RL}{reinforcement learning}
\acrodef{POCS}{Projections onto convex sets}
\acrodef{SINR}{signal-to-interference-ratio}
\acrodef{RNN}{recurrent neural network}
\acrodef{BS}{base station}
\acrodef{MISO}{multiple-input single-output}
\acrodef{MIMO}{multiple-input multiple-output}
\acrodef{mmWave}{millimeter wave}
\acrodef{RF}{radio frequency}
\acrodef{PS}{phase shifter}
\acrodef{TTD}{true time delayer}
\acrodef{FDD}{frequency-division duplex}
\acrodef{NN}{neural network}
\acrodef{GAN}{generative adversarial network}
\acrodefplural{GAN}{generative adversarial networks}
\acrodef{ULA}{uniform linear array}
\acrodef{BiCNN}{bi-directional convolutional neural network}
\acrodef{EDN}{encoder-decoder network}
\acrodef{ISI}{inter-user interference}
\acrodef{KD-EDN}{KD-based EDN}
\acrodefplural{KD-EDN}{KD-based EDNs}
\acrodef{DNN}{deep neural network}
\acrodefplural{DNN}{deep neural networks}
\acrodef{KD}{knowledge distillation}
\acrodef{AOA}{angle of arrival}
\acrodef{AOD}{angle of departure}
\acrodef{URLLC}{ultra-reliable low-latency communication}
\acrodef{L2O}{learning to optimize}
\acrodef{MRT}{maximum ratio transmission}
\acrodef{BRB}{branch-reduce-bound}
\acrodef{DL}{deep learning}
\acrodef{CNN}{convolutional neural network}
\acrodef{WMMSE}{weighted-MMSE}
\acrodef{GNN}{graphical neural network}
\acrodef{IRS}{intelligent reflect surface}
\acrodef{CS}{compressed sensing}
\acrodef{SVD}{singular value decomposition}
\acrodef{ADMM}{alternating direction method of multipliers}
\acrodef{Tanh}{hyperbolic tangent}
\acrodef{FCNN}{fully-connected neural network}
\acrodefplural{FCNN}{fully-connected neural networks}
\acrodef{STE}{straight-through estimator}
\acrodef{TTO}{test-time optimization}
\acrodef{SGD}{stochastic gradient descent}
\acrodef{NLoS}{none-line-of-sight}
\acrodef{LoS}{line-of-sight}
\acrodef{THz}{Terahertz}

\begin{document}

\title{An Encoder-Decoder Network for Beamforming over Sparse Large-Scale MIMO Channels}

\author{
\IEEEauthorblockN{Yubo Zhang, Jeremy Johnston, and Xiaodong Wang}\\
\thanks{Y.~Zhang and J.~Johnston and X.~Wang are with the Department of Electrical Engineering, Columbia University, New York, NY 10027. E-mails: \{yz4891, jj3057, xw2008\}@columbia.edu.}
}
\maketitle
\date{}

\begin{abstract}
\label{sec:Abstract}
We develop an end-to-end deep learning framework for downlink beamforming in large-scale sparse \ac{MIMO} channels. The core is a deep \ac{EDN} architecture with three modules: (i) an encoder \ac{NN}, deployed at each user end, that compresses estimated downlink channels into low-dimensional latent vectors. The latent vector from each user is compressed and then fed back to the BS. (ii) a beamformer decoder \ac{NN} at the \ac{BS} that maps recovered latent vectors to beamformers, and (iii) a channel decoder \ac{NN} at the \ac{BS} that reconstructs downlink channels from recovered latent vectors to further refine the beamformers. The training of \ac{EDN} leverages two key strategies: (a) semi-amortized learning, where the beamformer decoder \ac{NN} contains an analytical gradient ascent during both training and inference stages, and (b) knowledge distillation, where the loss function consists of a supervised term and an unsupervised term, and starting from supervised training with MMSE beamformers, over the epochs, the model training gradually shifts toward unsupervised using the sum-rate objective. The proposed \ac{EDN} beamforming framework is extended to both far-field and near-field hybrid beamforming scenarios. Extensive simulations validate its effectiveness under diverse network and channel conditions.

\end{abstract}

\begin{IEEEkeywords}
Large-scale MIMO, sparse channel, downlink beamforming, deep learning, encoder-decoder network, semi-amortized learning, knowledge distillation, hybrid beamforming, near-field. 
\end{IEEEkeywords}

\section{Introduction} \label{sec:intro}

The next-generation wireless communication systems are characterized by higher carrier frequencies and large-scale antenna arrays, thereby significantly enhancing data rates and reducing latency \cite{larsson2014massive}. Among various physical-layer advancements, real-time downlink beamforming, whereby the \ac{BS} updates its transmit beamformers as the downlink user channels vary, is a key enabler of capacity-approaching data transmission and has attracted substantial research interests \cite{hashemi2017out,li2025real}. Conventional iterative beamforming methods, such as the \ac{WMMSE} algorithm \cite{shi2011iteratively} and the \ac{BRB} algorithm \cite{bjornson2013optimal}, even though achieving near-optimal solutions, are impractical for large-scale \ac{MIMO} systems due to their prohibitively high computational complexities. On the other hand, low-complexity schemes, such as the \ac{MRT}-based beamforming \cite{lozano2006optimum} and the linear \ac{MMSE} beamforming \cite{bjornson2014optimal}, yield solutions that typically exhibit a large gap to the capacity in large-scale systems. In principle, beamformer design can be viewed as finding a deterministic mapping from the space of downlink channel vectors to the space of beamformer vectors. More recently, various machine learning tools have been explored to obtain such a mapping.


\vspace{-5pt}
\subsection{Deep Learning for Beamforming and Channel Estimation} \label{dl_bf_knowledge}

The \ac{DL} technique has shown remarkable success in tackling non-convex optimization problems owing to its powerful feature learning and representation capabilities \cite{hinton2006fast}, and has therefore been applied to approximate the channel-to-beamformer mapping in a data-driven manner \cite{xia2019deep}. The work in \cite{huang2019deep} employed a \ac{DNN} as an autoencoder to optimize the beamformers in a supervised manner, while \cite{li2019deep} similarly utilized supervised learning to jointly optimize the channel estimation and beamforming. Recognizing that labeled sample acquisition is prohibitively time-consuming, recent studies have shifted toward unsupervised beamforming learning methods. The work in \cite{hojatian2021unsupervised} proposed an RSSI-based unsupervised approach to designing beamformers for massive \ac{MIMO} systems. An unsupervised hybrid beamforming design via an autoencoder framework was further introduced in \cite{peken2020unsupervised}. Furthermore, several studies have explored learning-based beamforming without explicit \ac{CSI}. The work in \cite{alkhateeb2018deep} proposes learning beamforming directly from received pilot signals in highly-mobile single-user systems. The authors in \cite{jiang2021learning} propose a \ac{GNN} framework that learns beamformers directly from noisy pilot signals and user location information.





\vspace{-5pt}
\subsection{Semi-amortized Learning and Knowledge Distillation} \label{semi_amor_intro}

The cost of optimization can be amortized over the data distribution by shifting the computational burden from online optimization to offline training \cite{amos2023tutorial}. As a result, the predicted solution for any input sample can be rapidly obtained via a forward mapping \cite{jin2023solution}, a paradigm commonly referred to as fully-amortized learning. To further enhance the quality of predicted solutions, semi-amortized learning schemes have been proposed \cite{kim2018semi}, which integrate model-based domain knowledge into the data-driven learning process. A representative semi-amortized model augments the solution mapping with analytical gradient descent during training \cite{andrychowicz2016learning}.

On the other hand, knowledge distillation (KD) method has received considerable attention in recent years \cite{gou2021knowledge}. It encompasses various techniques to transfer knowledge from a large network to a more compact one, enabling deep models to be deployed effectively on devices with constrained computational resources. In the context of KD, the student model is initially guided to mimic the behavior of the teacher model and progressively assumes greater autonomy over the learning process \cite{kim2018paraphrasing}. A practical KD-based training strategy is provided by \cite{hinton2015distilling}, which optimizes a \ac{KD} loss function composed of two key components: a student loss computed based on ground truth labels and a distillation loss derived from the teacher's softened output. 
 


\subsection{Contributions and Outline} \label{contri_outline}
Existing \ac{DL}-based beamformer designs typically find a direct mapping from the channel state, either perfect or noisy, to the beamformer; and they do not scale well when the numbers of antennas and users become large. In this paper, we develop an end-to-end \ac{DL}-based beamforming solution for \ac{FDD} large-scale MIMO systems. The major contributions of this work are as follows:
\begin{itemize}
    \item Our design takes into account various factors in a realistic \ac{CSI} feedback mechanism in \ac{FDD} \ac{MIMO} systems, such as dimension reduction, quantization and dequantization, source encoding and decoding, and the feedback channel error. Specifically, the estimated downlink channel at each user end is first mapped to a low-dimensional latent vector by an encoder network; the latent vector is then quantized and encoded using some source code; the resulting bit stream is then fed back to the \ac{BS}, which recovers the latent vector by source decoding and dequantization. The recovered latent vectors from all users are then mapped to both beamformers and reconstructed channels through a beamformer decoder and a channel decoder, respectively. Compared with the existing approaches based on direct channel-to-beamformer mapping, this architecture not only reduces the neural network size and therefore scales well for large-scale MIMO systems, but also significantly lowers the feedback overhead in \ac{FDD} systems.
    \item We employ two key learning strategies to train the proposed \ac{EDN} to obtain a mapping that leads to high-rate beamformers. The first is the knowledge-distillation strategy to effectively mitigate the convergence to local optima during the early stage of training. Specifically, the student model first imitates the teacher's behavior (i.e., the MMSE beamformer) in a supervised way, and then gradually transitions to directly optimizing an unsupervised objective, i.e., the sum rate. The second strategy is semi-amortized learning that incorporates analytical gradient ascent steps to improve the beamformer given by the \ac{NN} mapping, in both training and inference stages. 
    \item We illustrate that the proposed \ac{EDN} beamforming framework is applicable to diverse beamforming scenarios. Specifically, we consider both single-cell and spatial-division systems for user deployment, and extend the scheme to hybrid beamforming in both far-field and near-field regimes. Simulation results demonstrate that the proposed method generalizes well across different scenarios and SNR conditions, consistently exhibiting high sum rate. 
\end{itemize}


\vspace{-2pt}

\section{System Descriptions and Problem Formulation} \label{sec:sysmodel}

In this section, we present the downlink beamforming system model and the sparse \ac{mmWave} channel model, and then formulate the FDD downlink beamformer design problem.  

\vspace{-6pt}

\subsection{System Description} \label{sys_model}
We consider a downlink \ac{MIMO} beamforming system, where a \ac{BS} equipped with $N$ antennas employs a \ac{ULA} with half-wavelength spacing to serve $K$ users. Each user is equipped with $M$ antennas, and user cooperation is not permitted. Assume that a frequency-division duplex (FDD) transmission mode is employed. Due to the lack of channel reciprocity in \ac{FDD} systems, downlink channel estimation is performed at the user side, while the beamformer design is carried out at the \ac{BS}. Denote by $\Bar{\bm{H}}_k \in \mathbb{C}^{M \times N}$ the downlink channel matrix from the \ac{BS} to user $k$. During downlink transmission, the \ac{BS} applies a beamforming matrix $\bm{W}_k \in \mathbb{C}^{N \times M}$ to transmit the data signal $\bm{x}_k \in \mathbb{C}^{M}$ intended for user $k$. The received signal at user $k$ is given by
\begin{align} \label{receive_mimo_full_digit}
\bm{y}_k = \Bar{\bm{H}}_k \bm{W}_k \bm{x}_k + \Bar{\bm{H}}_k \sum_{i \neq k} \bm{W}_i \bm{x}_i + \bm{n}_k,
\end{align}
where $\bm{n}_k \sim \mathcal{CN}(\bm{0},\sigma_k^2 \bm{I}_M)$ represent the additive noise at user $k$. For notational simplicity, we define the normalized channel matrix of user $k$ as $\bm{H}_k \triangleq \frac{\Bar{\bm{H}}_k}{\sigma_k^2}$. The overall normalized downlink channel matrix is then given by $\bm{H}=[\bm{H}_1,\dots,\bm{H}_K]$, and the overall beamforming matrix is denoted as $\bm{W}=[\bm{W}_1,\dots,\bm{W}_K] \in \mathbb{C}^{N \times KM}$. The achievable sum rate of the system is given by
\begin{align} \label{mimo_sr_exp}
R_\text{sum}(\bm{H},\bm{W})=\sum_{k=1}^K  \log_2 \lvert \bm{I}_M + \bm{\Sigma}_k^{-1} \bm{H}_k \bm{W}_k \bm{W}_k^H \bm{H}_k^H \rvert, 
\end{align}
where $\bm{\Sigma}_k = \bm{I}_M + \bm{H}_k \left( \sum_{i \neq k} \bm{W}_i \bm{W}_i^H \right) \bm{H}_k^H$ is the covariance matrix of the interference-plus-noise term for user $k$. 

For the special case of a \ac{MISO} system where each user is equipped with a single antenna, i.e., $M=1$, the channel vector between the \ac{BS} and user $k$ is denoted by $(\Bar{\bm{h}}_k)^T \in \mathbb{C}^{1 \times N}$. The \ac{BS} applies a beamforming vector $\bm{w}_k \in \mathbb{C}^{N \times 1}$ to transmit the data symbol $x_k \in \mathbb{C}$ for user $k$. Then the received signal at user $k$ is
\begin{align} \label{receive_miso_full_digit}
y_k = (\Bar{\bm{h}}_k)^H \sum_{i=1}^{K} \bm{w}_i x_i + n_k, 
\end{align}
where $n_k \sim \mathcal{CN}(0,\sigma_k^2)$ denotes the additive noise at user $k$. Similarly as before, denote the normalized channel vector of user $k$ as $\bm{h}_k \triangleq \frac{\Bar{\bm{h}}_k}{\sigma_k^2}$, and define the overall normalized channel matrix as $\bm{H}=[\bm{h}_1,\dots,\bm{h}_K]$, and the overall beamforming matrix as $\bm{W}=[\bm{w}_1,\dots,\bm{w}_K] \in \mathbb{C}^{N \times K}$. The achievable sum rate in this \ac{MISO} setting is given by
\begin{align} \label{sinr_rate_miso_full}
R_\text{sum}(\bm{H},\bm{W}) = \sum_{k=1}^K \log_2 \left(1+\frac{\left|\bm{h}_k^H \bm{w}_k\right|^2}{1+\sum_{i \neq k} \left|\bm{h}_k^H \bm{w}_i\right|^2}\right).
\end{align}


\subsection{Sparse Channel Model} \label{sparse_chan}
We next introduce the \ac{mmWave} channel model, whose sparsity facilitates efficient channel compression. Specifically, we consider a geometric channel model with $L$ scattering paths, and the channel between the \ac{BS} and user $k$ is: 
\begin{align}  \label{mmwave_chan_user_k}
\Bar{\bm{H}}_k = \sqrt{\frac{MN}{L}} \sum_{\ell=1}^{L} \alpha_{k,\ell} \bm{a}_M \left(\beta_{k,\ell}, \frac{d_k}{\lambda} \right) \bm{a}^H_N \left(\gamma_{k,\ell},\frac{d}{\lambda}\right),
\end{align}
where $\alpha_{k,\ell} \sim \mathcal{CN}(0,1)$, $\beta_{k,\ell} \in [-\frac{\pi}{2},\frac{\pi}{2}]$, $\gamma_{k,\ell} \in [-\frac{\pi}{2},\frac{\pi}{2}]$ denote the complex gain, the \ac{AOA} and the \ac{AOD} of the $\ell^{\text{th}}$ scattering path between the \ac{BS} and user $k$, respectively; $d_k$ and $d$ denote the antenna spacings for user $k$ and the \ac{BS}, respectively; $\lambda$ is the wavelength. The function $\bm{a}_n(\cdot,\cdot) \in \mathbb{C}^n$ denotes the array response vector of a \ac{ULA} with $n$ antenna elements, given by:
\begin{align} \label{array_response_vec}
\bm{a}_n \left( \theta,\frac{d}{\lambda} \right) = \frac{1}{\sqrt{n}} \left[ 1,e^{j \frac{2 \pi d}{\lambda} \sin{\theta}}, \dots, e^{j \frac{2 \pi d}{\lambda} (n-1) \sin{\theta}} \right]^T,
\end{align}
where $\theta$, $d$ and $\lambda$ denote the steering angle, antenna spacing and wavelength, respectively. Throughout this paper, half-wavelength antenna spacing is assumed, i.e., $d_k = d = \frac{\lambda}{2}$, so that $\bm{a}_n \left(\theta, \frac{1}{2}\right) = \frac{1}{\sqrt{n}} \left[ 1,e^{j \pi \sin{\theta}}, \dots, e^{j \pi (n-1) \sin{\theta}} \right]^T$. The constant $\sqrt{\frac{MN}{L}}$ in \eqref{mmwave_chan_user_k} leads to $\mathbb{E}[\|\Bar{\bm{H}}_k\|_F^2]=MN$. Note that for the \ac{MISO} case where $M=1$, the receive array response degenerates to the scalar form $\bm{a}_1(\cdot,\cdot) = 1$.

\begin{figure*}
     \centering
     \justifying
     \begin{subfigure}[b]{0.48\textwidth}
         \centering
         \includegraphics[width=6.1cm]{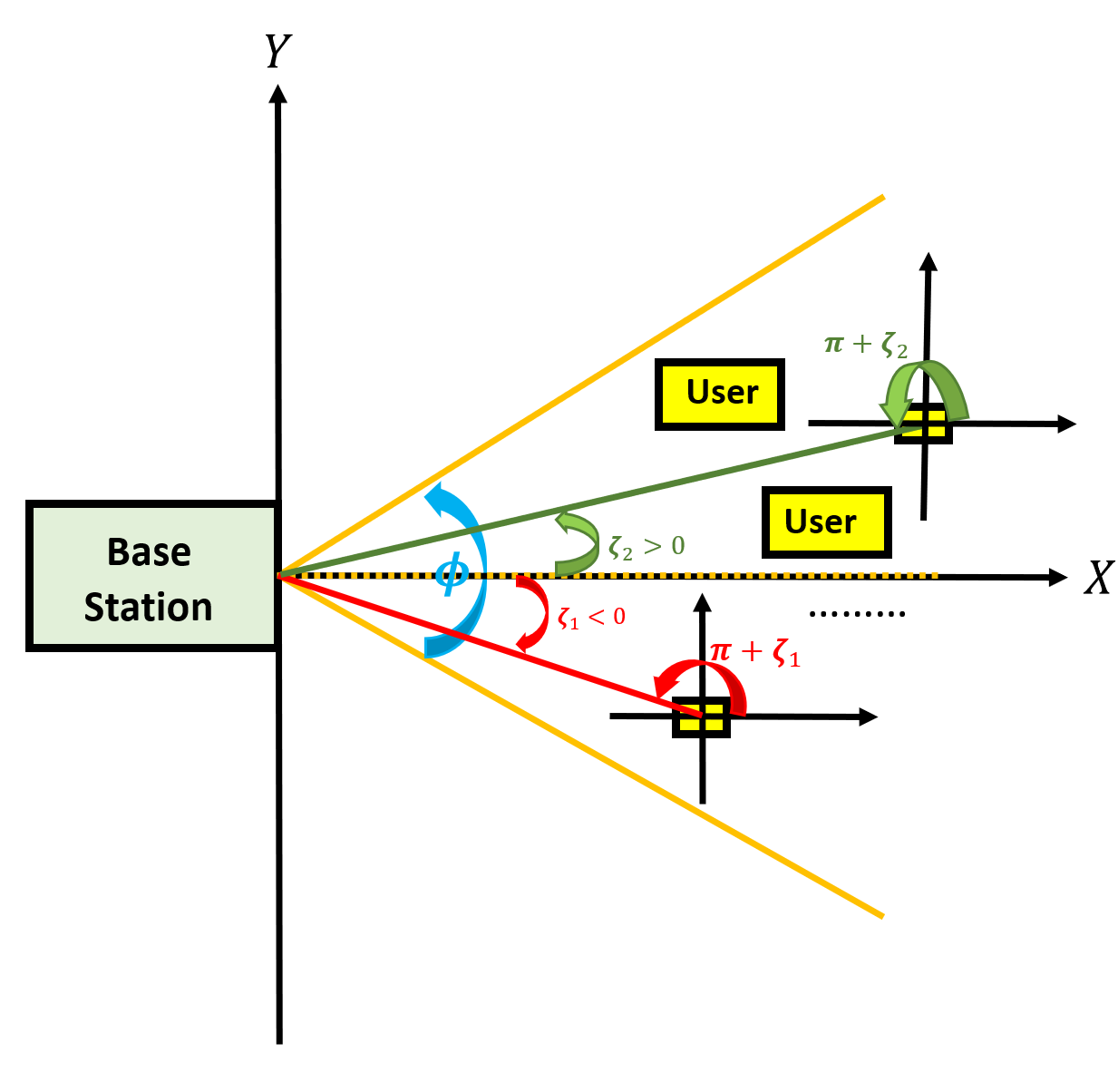}
            \caption{Single-cell system.}
		\label{single_cell_sys}
     \end{subfigure}
     \begin{subfigure}[b]{0.48\textwidth}
        \centering
        \includegraphics[width=6.0cm]{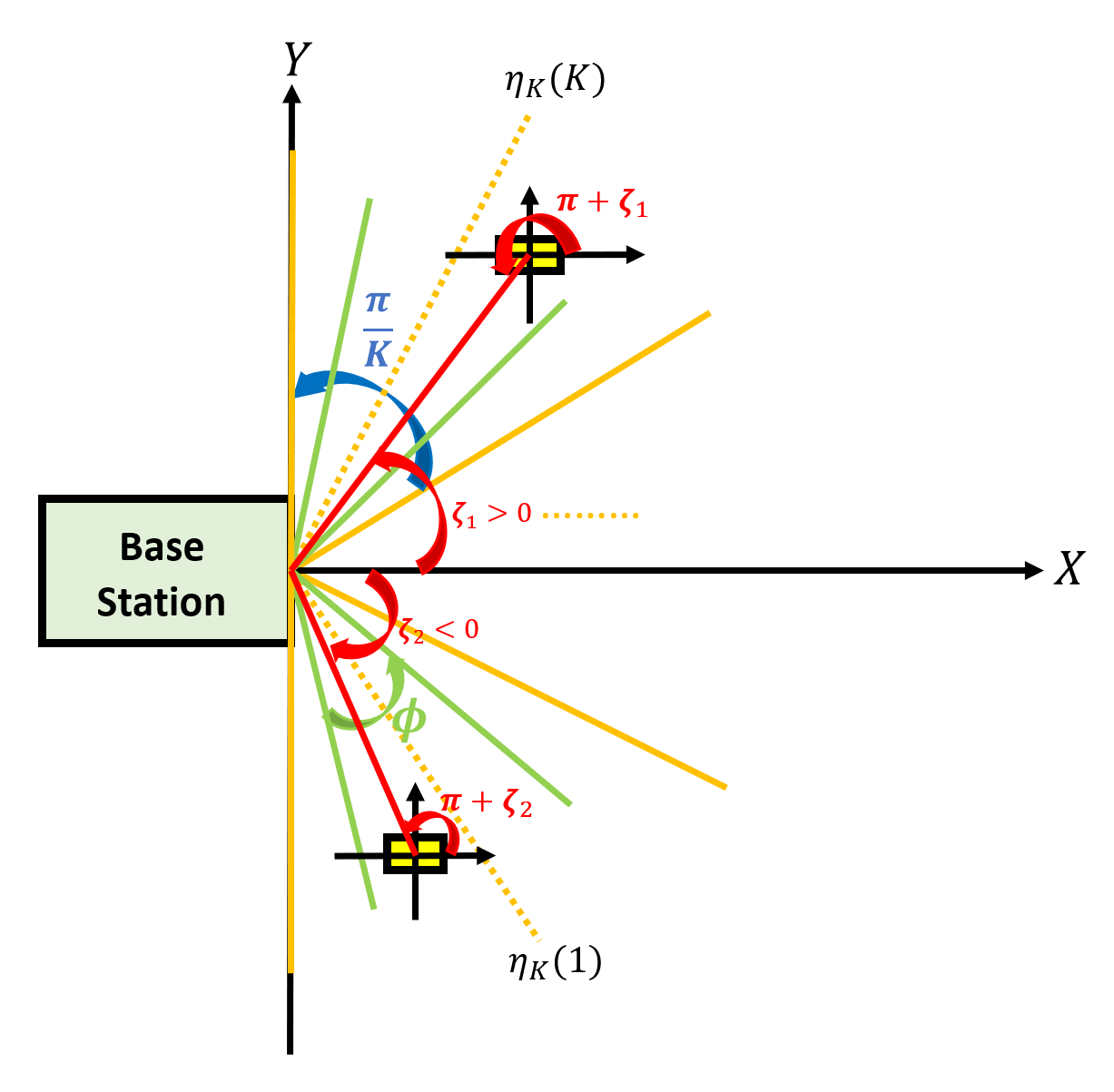}
            \caption{Spatial-division system.}
	    \label{spatial_division_sys}
    \end{subfigure}
        \caption{Single-cell and spatial-division downlink systems.}
        \label{sys_graph}
\end{figure*}

In this paper, we consider two geographical distribution scenarios of the users, as illustrated in Fig.~\ref{sys_graph}. In a single-cell system with a sector angle $\phi \in [0,\pi]$, all $K$ users are randomly located in the angular range $[-\frac{\phi}{2},\frac{\phi}{2}]$. On the other hand, in a spatial-division system, the angular range $[-\frac{\pi}{2},\frac{\pi}{2}]$ is first evenly divided into $K$ sectors, with the center angle of the $k^{\text{th}}$ sector given by $\eta_K(k) = -\frac{\pi}{2} + \frac{(k-\frac{1}{2})\pi}{K}$. User $k$ is randomly located in the range $[\eta_K(k)-\frac{\psi}{2},\eta_K(k)+\frac{\psi}{2}]$, where $\psi \in [0,\frac{\pi}{K}]$ and $k \in [K]$.    
We assume that each user is located in either the first or the fourth quadrant and oriented directly toward the \ac{BS}. As shown in Fig.~\ref{sys_graph}, for each user $k$ with azimuth angle $\zeta_k \in [-\frac{\pi}{2},\frac{\pi}{2}]$, the mean \ac{AOD} and \ac{AOA} are defined as $\Bar{\gamma}_k = \zeta_k, \ \Bar{\beta}_k = \pi + \zeta_k$, $k \in [K]$. This relationship applies to both the single-cell system with $\zeta_k \in [-\frac{\phi}{2},\frac{\phi}{2}]$ and spatial-division system with $\zeta_k \in [\eta_K(k)-\frac{\psi}{2},\eta_K(k)+\frac{\psi}{2}]$. To account for the angular dispersion introduced by multi-path propagation, a normal distribution is used to generate the individual path angles. Specifically, the AOD and AOA of the $\ell^{\text{th}}$ path for user $k$ are modeled as
\begin{align} \label{multi_ang_calc}
\gamma_{k,\ell} \sim \mathcal{N}(\Bar{\gamma}_k, \sigma_{\gamma,k}^2), \
\beta_{k,\ell} \sim \mathcal{N}(\Bar{\beta}_k, \sigma_{\beta,k}^2),
\end{align}
where $\ell \in[L], k \in [K]$, $\sigma_{\gamma,k}$ and $\sigma_{\beta,k}$ denote the standard deviations of the departure and arrival angles, respectively. As a result, the downlink channel $\Bar{\bm{H}}_k$ for user $k$ can be constructed according to \eqref{mmwave_chan_user_k}-\eqref{multi_ang_calc} and normalized as $\bm{H}_k = \frac{\Bar{\bm{H}}_k}{\sigma_k^2}$. 

\vspace{-5pt}

\subsection{Problem Formulation} \label{prob_formu}

For the downlink transmission, each channel coherence interval is divided into the pilot phase and the data transmission phase. During the pilot phase, the \ac{BS} transmits pilot symbols to mobile users. Each user then performs channel estimation based on its received pilot signal and feeds back the estimated channel to the \ac{BS}. The BS then computes the downlink beamformers for all users based on received estimated channels, which are employed during the data transmission phase. 

Let $\Tilde{\bm{H}}_k$ denote the estimated normalized channel matrix for user $k$. As discussed in Sec.~\ref{sparse_chan}, the channel exhibits inherent sparsity, making it compressible. To exploit this, each estimated channel $\Tilde{\bm{H}}_k$ undergoes a dimension reduction operation via a mapping $g(\cdot)$, i.e., $\bm{z}_k = g(\Tilde{\bm{H}}_k)$, where $\bm{z}_k \in \mathbb{R}^{d \times 1}$ is a low-dimensional real-valued latent vector with $d \ll 2 M N$. Each user $k$ then quantizes $\bm{z}_k$, compresses the resulting binary sequence using a lossless source coding scheme, e.g., entropy coding, and transmits the compressed bits to the \ac{BS} over a feedback channel. The \ac{BS} performs source decoding followed by dequantization to reconstruct the latent vector $\hat{\bm{z}}_k$ for each user $k$. Denote $\hat{\bm{z}}=[\hat{\bm{z}}_1,\dots,\hat{\bm{z}}_K]$. The beamforming matrix $\bm{W}=[\bm{W}_1,\dots,\bm{W}_K] \in \mathbb{C}^{N \times KM}$ is then computed based on the reconstructed latent vector via a mapping $f(\cdot)$, i.e., $\bm{W} = f(\hat{\bm{z}})$. Our objective is to jointly design the mappings $g: \Tilde{\bm{H}}_k \rightarrow \bm{z}_k$ and $f: \hat{\bm{z}} \rightarrow \bm{W}$, such that the average downlink sum rate is maximized, i.e., 
\begin{align} \label{unified_formulation}
\max_{\substack{\bm{z}_k = g(\Tilde{\bm{H}}_k), \\ \bm{W} = f(\hat{\bm{z}})}} \mathbb{E}_{\bm{H},\Delta_{\bm{H}},\Delta \bm{z}} \left[R_{\text{sum}}(\bm{H},\bm{W})\right], \text{s.t.} \ \| \bm{W}\|_{\text{F}}^2 \leq P.
\end{align}
Here, $\Delta_{\bm{H}}=\Tilde{\bm{H}}_k-\bm{H}_k$ denotes the channel estimation error between the true channel $\bm{H}_k$ and its estimate $\Tilde{\bm{H}}_k$, $\Delta \bm{z}=\hat{\bm{z}}_k-\bm{z}_k$ is the discrepancy between the true latent vector $\bm{z}_k$ and the reconstructed latent vector $\hat{\bm{z}}_k$ at the \ac{BS}, which includes both the quantization error and the feedback channel distortion. The achievable sum rate is then computed using \eqref{mimo_sr_exp} for the \ac{MIMO} case and \eqref{sinr_rate_miso_full} for the \ac{MISO} case. 

The formulation in \eqref{unified_formulation} is a stochastic functional optimization for which there is no known conventional solver. In this paper, we represent the mapping functions $g(\cdot)$ and $f(\cdot)$ in \eqref{unified_formulation} using \acp{DNN}, and propose effective data-driven learning schemes to solve \eqref{unified_formulation}.

\section{Deep Encoder-Decoder Network for Beamforming} \label{DL_framework}

\subsection{Network Architecture} \label{net_arch_intro}

\begin{figure*}
    \centering
    \includegraphics[width=0.9\linewidth]{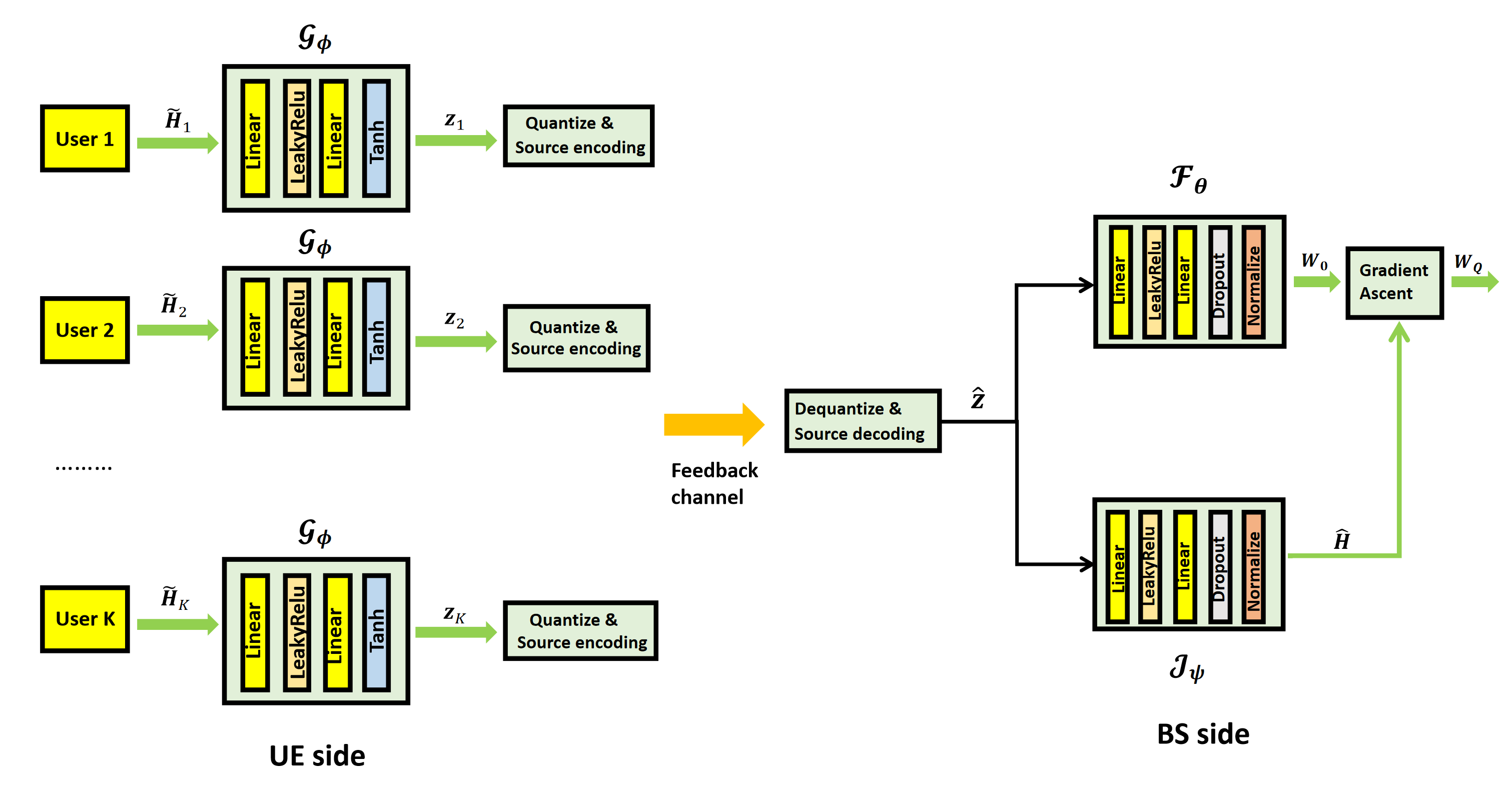}
    \caption{The architecture of the encoder-decoder neural network (NN) for downlink beamforming.}
    \label{edn_architecture}
\end{figure*}

We now design a deep encoder-decoder network (EDN) tailored to the problem in \eqref{unified_formulation}. The proposed \ac{EDN}, depicted in Fig.~\ref{edn_architecture}, consists of three subnetworks: (a) a shared encoder \ac{NN} $\mathcal{G}_{\bm{\phi}}$ which approximates the mapping $g(\cdot)$ in \eqref{unified_formulation} and is utilized by all users, with learnable parameters $\bm{\phi}$; (b) a beamformer decoder \ac{NN} $\mathcal{F}_{\bm{\theta}}$ which approximates the mapping $f(\cdot)$ in \eqref{unified_formulation} at the \ac{BS}, with learnable parameters $\bm{\theta}$; and (c) a channel decoder \ac{NN} $\mathcal{J}_{\bm{\psi}}$ also at the \ac{BS}, with learnable parameters $\bm{\psi}$.

\subsubsection{Encoder} \label{enc_structure} 

The large-scale \ac{MIMO} channels in high-frequency bands, e.g., the mmWave and terahertz bands, are inherently sparse due to limited scatterings \cite{alkhateeb2014channel}. Such channel sparsity has been extensively exploited in \ac{MIMO} communications. In particular, \cite{balevi2020high} employs a \ac{GAN} to approximate the mapping from low-dimensional latent vectors to high-dimensional sparse channels. Motivated by this line of work, we also compress the sparse channels using an encoder \ac{NN} prior to beamforming optimization. Consequently, the target channel-to-beamformer mapping is decomposed into a channel-to-latent mapping followed by a latent-to-beamformer mapping, which offers two key advantages. First, in \ac{FDD} downlink systems, each user only feeds back a low-dimensional latent vector rather than the full channel estimate to the \ac{BS}, thereby reducing the feedback overhead. Secondly, as noted in \cite{johnston2023curriculum}, directly learning a channel-to-beamformer mapping in large-scale systems is difficult, since training often converges to suboptimal solutions due to the large neural network size. Intuitively, introducing latent representations can mitigate this issue by enabling a more compact neural network with an improved convergence behavior. Notably, we employ a DNN rather than a GAN for channel compression and reconstruction in this work, as \acp{GAN} are prone to mode collapse and vanishing gradient issues when trained on high-dimensional data samples \cite{wiatrak2019stabilizing}. 

\vspace{-1pt}
Specifically, the input to the encoder \ac{NN} at user $k$ is an estimate of its normalized channel $\bm{H}_k$, denoted as $\Tilde{\bm{H}}_k$, obtained based on the received pilot signals. Denote the real-valued input vector to the encoder $\mathcal{G}_{\bm{\phi}}$ as $\Tilde{\bm{h}}_k = \left[ \text{vec}\left( \Re\{\Tilde{\bm{H}}_k\} \right), \text{vec}\left( \Im\{\Tilde{\bm{H}}_k\} \right) \right] \in \mathbb{R}^{2MN}$. $\mathcal{G}_{\bm{\phi}}$ then outputs a compact latent representation $\bm{z}_k \in \mathbb{R}^{d \times 1}$. As illustrated in Fig.~\ref{edn_architecture}, the encoder $\mathcal{G}_{\bm{\phi}}$ consists of several fully connected linear layers, with a LeakyReLU activation layer between two adjacent linear layers to enhance the network’s expressivity and to mitigate the vanishing gradient problem. The final layer is composed of \ac{Tanh} activation functions, so that each element of the output latent vector $\bm{z}_k$ is in the range $[-1,1]$ in order to facilitate subsequent quantization. 



\subsubsection{Channel Decoder} \label{chan_dec_structure}
Each user channel is individually recovered by a shared channel decoder \ac{NN} $\mathcal{J}_{\bm{\psi}}$ at the \ac{BS}. Specifically, for user $k$, the input to $\mathcal{J}_{\bm{\psi}}$ is the recovered latent vector $\bm{\hat{z}}_k$, and the output is the vectorized recovered normalized channel $\bm{\hat{h}}_k \in \mathbb{R}^{2MN}$. The network $\mathcal{J}_{\bm{\psi}}$ consists of fully-connected linear layers, activation layers, and batch normalization layers to enhance training stability. The output vector $\hat{\bm{h}}_k \in \mathbb{R}^{2MN}$ is then reshaped as $\hat{\bm{H}}_k = \text{vec}_h^{-1} \left(\hat{\bm{h}}_k \left[1 : MN\right]\right) + j \cdot \text{vec}_h^{-1} \left(\hat{\bm{h}}_k \left[MN+1 : 2MN\right]\right)$, where $\text{vec}_h^{-1}$ reshapes an $MN$ vector into an $M \times N$ matrix. The overall reconstructed normalized channel is $\hat{\bm{H}} = [\hat{\bm{H}}_1,\dots,\hat{\bm{H}}_K]$.

\subsubsection{Beamformer Decoder} \label{bf_dec_structure}
The input to the beamformer decoder \ac{NN} $\mathcal{F}_{\bm{\theta}}$ is the reconstructed latent vector $\bm{\hat{z}} = [\bm{\hat{z}}_1,\dots,\bm{\hat{z}}_K]$, and its output is the vectorized overall beamformer $\bm{w} \in \mathbb{R}^{2NKM \times 1}$. The network $\mathcal{F}_{\bm{\theta}}$ consists of several linear layers, with LeakyReLU activation functions inbetween. To further enhance robustness under varying channel conditions and to reduce overfitting, a dropout layer with a fixed dropout rate of 0.2 is incorporated after each activation layer. A final normalization layer scales the output beamformer to meet the total power constraint $P$, such that $\bm{w} = \sqrt{P} \cdot \frac{\bm{\Tilde{\bm{w}}}}{\| \bm{\Tilde{\bm{w}}} \|_2}$, 
where $\bm{\Tilde{\bm{w}}}$ is the output from the last linear layer. The output vector $\bm{w} \in \mathbb{R}^{2NKM \times 1}$ is then reshaped to form a complex-valued beamforming matrix $\bm{W}_0 = \text{vec}_w^{-1} \left(\bm{w} \left[1 : NKM \right]\right) + j \cdot \text{vec}_w^{-1} \left(\bm{w} \left[NKM+1 : 2NKM \right]\right)$, where $\text{vec}_w^{-1}$ denotes the inverse vectorization operation that reshapes a real-valued vector of size $NKM \times 1$ into a matrix of size $N \times KM$. The final step is known as semi-amortized optimization \cite{amos2023tutorial}. Specifically, as in Fig.~\ref{edn_architecture}, starting from $\bm{W}_0$ and using the reconstructed normalized channel $\hat{\bm{H}}$, we perform $Q_i$ steps of gradient ascent to further increase the sum rate $R_{\text{sum}}(\bm{H},\bm{W})$ given in \eqref{mimo_sr_exp} for the \ac{MIMO} case and \eqref{sinr_rate_miso_full} for the \ac{MISO} case, to obtain the final beamformer $\bm{W}_{Q_i}$:
\begin{align} \label{gd_part_inference}
\bm{W}_q = \bm{W}_{q-1} + \eta \cdot \nabla_{\bm{W}} R_{\text{sum}}(\hat{\bm{H}},\bm{W}_{q-1}), \ q \in [Q_i].
\end{align}
The encoder $\mathcal{G}_{\bm{\phi}}$ and the beamformer decoder $\mathcal{F}_{\bm{\theta}}$ are jointly trained first, which is described in Sec. III-B. The channel decoder $\mathcal{J}_{\bm{\psi}}$ is then trained given the trained encoder, which is described in Sec.~\ref{chan_dec_train}.

\vspace{-6pt}
\subsection{Beamforming EDN Training} \label{train_design_edn}

\subsubsection{Semi-Amortized Optimization and Alternating Training} \label{semi_amor_alter_train}
During the \ac{EDN} training phase, noise-free normalized channel matrices $\bm{H}_k$ are randomly generated based on the \ac{mmWave} channel model described in Sec.~\ref{sparse_chan}. Associated with each sample of $\bm{H}_k$, there are two error samples: the channel estimation error $\Delta \bm{H}_k \triangleq \Tilde{\bm{H}_k} - \bm{H}_k$, and the feedback error $\Delta \bm{z}_k \triangleq \hat{\bm{z}}_k - \bm{z}_k$. Therefore in each training epoch the training data batch has the form $\{(\bm{H}^{(j)},\Delta \bm{H}^{(j)}, \Delta \bm{z}^{(j)})\}_{j=1}^{N_{\text{b}}}$, where $N_{\text{b}}$ is the batch size, $\Delta \bm{H} \triangleq [\Delta \bm{H}_1,\dots,\Delta \bm{H}_K]$ and $\Delta \bm{z} \triangleq [\Delta \bm{z}_1,\dots,\Delta \bm{z}_K]$. 



Recall Sec.~\ref{semi_amor_intro}, we adopt a semi-amortized optimization framework that integrates domain knowledge into the learning process to solve \eqref{unified_formulation}. The objective is to facilitate the learning of the mappings $\mathcal{F}_{\bm{\theta}}(\cdot)$ and $\mathcal{G}_{\bm{\phi}}(\cdot)$, such that they generate a high-quality initial beamformer $\bm{W}_0$ that enables convergence to a near-optimal beamformer $\bm{W}_{Q}$ within $Q$ steps of gradient ascent. To that end, we define the loss function 
\begin{align} \label{origin_loss_fun}
L(\bm{\phi},\bm{\theta}) = - \mathbb{E}_{\bm{H}, \Delta \bm{H}, \Delta \bm{z}} \{ R_{\text{sum}}(\bm{H},\omega(\bm{H},\Delta \bm{H}, \Delta \bm{z})) \}, 
\end{align}
where the beamformer $\omega(\bm{H},\Delta \bm{H}, \Delta \bm{z})$ is computed as follows: first we compute the recovered latent variables of all users at the \ac{BS}, denoted by $\hat{\bm{z}}=[\hat{\bm{z}}_1,\dots,\hat{\bm{z}}_K]$, where $\hat{\bm{z}}_k = \mathcal{G}_{\bm{\phi}} (\Tilde{\bm{H}}_k) + \Delta \bm{z}_k$, $k\in[K]$. Then using the beamformer decoder $\mathcal{F}_{\bm{\theta}}$, we obtain the initial beamformer $\bm{W}_0 = \mathcal{F}_{\bm{\theta}}(\hat{\bm{z}})$. We further refine the beamformer by performing $Q_t$ steps of gradient ascent as follows,
\begin{align} \label{gd_part_train}
\bm{W}_q = \bm{W}_{q-1} + \eta \cdot \nabla_{\bm{W}} R_{\text{sum}}(\bm{H},\bm{W}_{q-1}), \ q \in [Q_t].
\end{align}
Note that different from \eqref{gd_part_inference} for the inference stage, where the recovered channel $\hat{\bm{H}}$ is used, during the training stage, in \eqref{gd_part_train}, the true channel $\bm{H}$ is used.

Finally we set $\omega(\bm{H},\Delta \bm{H}, \Delta \bm{z})=\bm{W}_{Q_t}$. The semi-amortized optimization then learns the parameters of the \ac{EDN} through the analytical gradient steps in \eqref{gd_part_train}.

Denoting $\bm{w} = \text{vec}(\bm{W})$ and $\Bar{\bm{w}}$ as the complex conjugate of $\bm{w}$. According to the chain rule of differentiation for complex-valued variables, the gradient of the loss function in \eqref{origin_loss_fun} with respect to the network parameters $\bm{\theta}$ is given as
\begin{align} \label{theta_partial}
&\nabla_{\bm{\theta}} L(\bm{\phi},\bm{\theta}) = \mathbb{E}_{\bm{H},\Tilde{\bm{H}}} \left\{ \frac{\partial L}{\partial \bm{w}_0} \cdot \frac{\partial \bm{w}_0}{\partial \bm{\theta}} + \frac{\partial L}{\partial \Bar{\bm{w}}_0} \cdot \frac{\partial \Bar{\bm{w}}_0}{\partial \bm{\theta}} \right\} \notag \\
&= \mathbb{E}_{\bm{H},\Tilde{\bm{H}}}  \left( \frac{\partial L}{\partial \bm{w}_Q} \frac{\partial \bm{w}_Q}{\partial \bm{w}_0} + \frac{\partial L}{\partial \Bar{\bm{w}}_Q} \frac{\partial \Bar{\bm{w}}_Q}{\partial \bm{w}_0} \right)  \cdot \frac{\partial \bm{w}_0}{\partial \bm{\theta}} \notag \\
&+ \left( \frac{\partial L}{\partial \bm{w}_Q} \frac{\partial \bm{w}_Q}{\partial \Bar{\bm{w}}_0} + \frac{\partial L}{\partial \Bar{\bm{w}}_Q} \frac{\partial \Bar{\bm{w}}_Q}{\partial \Bar{\bm{w}}_0} \right)  \cdot \frac{\partial \Bar{\bm{w}}_0}{\partial \bm{\theta}} .
\end{align}
Note that the gradient terms $\frac{\partial L}{\partial \bm{w}_Q} = - \frac{\partial R_{\text{sum}}}{\partial \bm{w}_Q}$ and $\frac{\partial L}{\partial \Bar{\bm{w}}_Q} = - \frac{\partial R_{\text{sum}}}{\partial \Bar{\bm{w}}_Q}$ can be computed analytically from the closed-form expression of sum-rate objective. Moreover, $\frac{\partial \bm{w}_0}{\partial \bm{\theta}}$ and $\frac{\partial \Bar{\bm{w}}_0}{\partial \bm{\theta}}$ represent the gradients of neural network outputs with respect to its parameters, which are obtained via automatic differentiation. The intermediate gradient terms $\frac{\partial \bm{w}_Q}{\partial \bm{w}_0}$, $\frac{\partial \Bar{\bm{w}}_Q}{\partial \bm{w}_0}$, $\frac{\partial \bm{w}_Q}{\partial \Bar{\bm{w}}_0}$ and $\frac{\partial \Bar{\bm{w}}_Q}{\partial \Bar{\bm{w}}_0}$ are computed based on \eqref{gd_part_train} and given in the Appendix. The gradient $\nabla_{\bm{\phi}} L(\bm{\phi},\bm{\theta})$ can be similarly computed.


An alternating optimization scheme is employed to train $\mathcal{G}_{\bm{\phi}}$ and $\mathcal{F}_{\bm{\theta}}$ jointly based on \eqref{theta_partial}. In each training epoch, the encoder $\mathcal{G}_{\bm{\phi}}$ is first updated by $N_{\text{en}}$ \ac{SGD} steps, after which the beamformer decoder $\mathcal{F}_{\bm{\theta}}$ is updated by $N_{\text{de}}$ \ac{SGD} steps, both using the training data batch $\{(\bm{H}^{(j)}, \Delta \bm{H}^{(j)}, \Delta \bm{z}^{(j)})\}_{j=1}^{N_b}$. Denote $\bm{\phi}^{(n,0)}=\bm{\phi}^{(n)}$, $\bm{\theta}^{(n,0)}=\bm{\theta}^{(n)}$, $\bm{\phi}^{(n,N_{\text{en}})}=\bm{\phi}^{(n+1)}$, and $\bm{\theta}^{(n,N_{\text{de}})}=\bm{\theta}^{(n+1)}$, for $n=0,1,\dots$. The parameter updates during the $n^{\text{th}}$ training epoch are as follows:
\begin{subequations} \label{update_alter}
\begin{align} 
\label{update_alter_en}
&\bm{\phi}^{(n,i)} = \bm{\phi}^{(n,i-1)} - \eta \nabla_{\bm{\phi}}L(\bm{\phi}^{(n,i-1)},\bm{\theta}^{(n)}), i \in [N_{\text{en}}], \\
\label{update_alter_de}
&\bm{\theta}^{(n,i)} = \bm{\theta}^{(n,i-1)} - \eta \nabla_{\bm{\theta}} L(\bm{\phi}^{(n+1)},\bm{\theta}^{(n,i-1)}), i \in [N_{\text{de}}]. 
\end{align}  
\end{subequations}
Note that since the encoder network $\mathcal{G}_{\bm{\phi}}$ is smaller than the beamformer decoder network $\mathcal{F}_{\bm{\theta}}$, $\mathcal{G}_{\bm{\phi}}$ in general requires less \ac{SGD} steps than $\mathcal{F}_{\bm{\theta}}$, i.e., $N_{\text{en}} < N_{\text{de}}$. 



\subsubsection{Enhanced Training with Knowledge Distillation} \label{kd_enhance_train}



Considering the limited expressive power of the fully-connected neural networks, it is challenging to train the \ac{EDN} only based on the non-convex sum-rate objective \cite{johnston2023curriculum}. To address this issue, inspired by the \ac{KD} method introduced in Sec.~\ref{semi_amor_intro}, we employ high-quality supervisions at the early stage of training, which help the model rapidly learn a good solution mapping first, and then the learning process gradually transitions to the original unsupervised sum-rate objective, allowing the network to fine-tune its parameters for improved performance. To implement this strategy, a hybrid training model, called \ac{KD-EDN}, is proposed. Similar to \cite{hinton2015distilling}, we define a \ac{KD} loss function, which consists of a supervised term $L^{\text{s}}(\bm{\phi},\bm{\theta})$ and an unsupervised term $L^{\text{u}}(\bm{\phi},\bm{\theta})$, to train the \ac{KD-EDN}. In the early training stages, we prioritize guiding the \ac{KD-EDN} to closely mimic the teacher, enabling it to learn the underlying channel distributions through high-quality beamformers. As training progresses, the \ac{KD-EDN} gradually acquires the capacity to learn independently. To facilitate this transition, we gradually shift the training objective from teacher supervision to unsupervised optimization by adjusting the weights of $L^{\text{s}}$ and $L^{\text{u}}$. The KD loss function is defined as:
\begin{align} \label{proposed_kd_loss_func}
L_{\alpha}(\bm{\phi},\bm{\theta}) = \alpha L^{\text{u}}(\bm{\phi},\bm{\theta}) + (1-\alpha) L^{\text{s}}(\bm{\phi},\bm{\theta}), \ \alpha \in [0,1],
\end{align}
where the weight $\alpha$ is gradually increased from 0 to 1 over training. The unsupervised loss is the sum rate in \eqref{origin_loss_fun}, i.e., 
\begin{align} \label{rate_loss_goal}
L^{\text{u}} = - \mathbb{E}_{\bm{H}, \Delta \bm{H}, \Delta \bm{z}} \left\{ R_{\text{sum}}(\bm{H},\omega(\bm{H},\Delta \bm{H}, \Delta \bm{z})) \right\},
\end{align}
where $\omega(\bm{H},\Delta \bm{H}, \Delta \bm{z})=\bm{W}_{Q_t}$ is given in \eqref{gd_part_train}. We assume that the teacher's knowledge corresponding to the true channel realization $\bm{H} = [\bm{H}_1,\dots,\bm{H}_K]$ is given by its \ac{MMSE} beamformer solution, denoted by $\Tilde{\bm{W}} = [\Tilde{\bm{W}}_1,\dots,\Tilde{\bm{W}}_K]$. To compute the \ac{MMSE} beamformer, define $\bm{M} = (\bm{I}_N + \sum_{i=1}^{K} \frac{P}{K} \bm{H}_i^H \bm{H}_i)^{-1}$, and we have $\Tilde{\bm{W}}_k = \sqrt{\frac{P}{K}} \cdot \frac{ \bm{M} \bm{H}_k^H}{ \Vert \bm{M} \bm{H}_k^H \Vert_2}$, for $k \in [K]$. Then the supervised loss is defined as 
\begin{align} \label{mse_loss_goal}
L^{\text{s}} = \mathbb{E}_{\bm{H},\Delta \bm{H}, \Delta \bm{z}} \left\{ \| \Tilde{\bm{W}} - \omega(\bm{H},\Delta \bm{H}, \Delta \bm{z}) \|_F^2 \right\}.
\end{align}
The gradient of the \ac{KD} loss function in \eqref{proposed_kd_loss_func} with respect to the parameters $\bm{\theta}$ or $\bm{\phi}$ is also given by \eqref{theta_partial}, with the terms $\frac{\partial L}{\partial \bm{w}_Q} = -\alpha \cdot \frac{\partial R_{\text{sum}}}{\partial \bm{w}_Q} + (1-\alpha) \cdot \frac{\partial \| \Tilde{\bm{W}} - \bm{W}_Q \|_F^2}{\partial \bm{w}_Q}$ and $\frac{\partial L}{\partial \Bar{\bm{w}}_Q} = -\alpha \cdot \frac{\partial R_{\text{sum}}}{\partial \Bar{\bm{w}}_Q} + (1-\alpha) \cdot \frac{\partial \| \Tilde{\bm{W}} - \bm{W}_Q \|_F^2}{\partial \Bar{\bm{w}}_Q}$.

\subsection{Channel Decoder Training} \label{chan_dec_train}
The channel decoder $\mathcal{J}_{\bm{\psi}}$ is trained after the encoder and the beamformer decoder are jointly trained, aiming to output an accurate channel estimate based on the recovered latent vector. In particular, it is trained in a supervised manner, based on the trained encoder $\mathcal{G}_{\bm{\phi}^*}$. Given the training data batch of each user $k$, denoted as $\{(\bm{H}_k^{(j)},\Delta \bm{H}_k^{(j)}, \Delta \bm{z}_k^{(j)})\}_{j=1}^{N_{\text{b}}}$, for $k\in[K]$, the channel decoder loss function is given by 
\begin{multline} \label{recovery_formu}
\text{$L(\psi) = \sum_{k=1}^{K} \mathbb{E}_{\bm{H}_k,\bm{H}_k, \Delta \bm{z}_k} \bigg\{ \| \bm{H}_k - $} \\
\text{$\mathcal{J}_{\bm{\psi}}(\mathcal{G}_{\bm{\phi}^*}(\bm{H}_k + \Delta \bm{H}_k) + \Delta \bm{z}_k) \|_F^2 \bigg\}$}
\end{multline} 
Note that the trained channel decoder $\mathcal{J}_{\bm{\psi}^*}$ is used during the inference stage, in the final gradient ascent step \eqref{gd_part_inference}, where $\hat{\bm{H}} = [\hat{\bm{H}}_1,\dots,\hat{\bm{H}}_K]$, with $\hat{\bm{H}}_k = \mathcal{J}_{\bm{\psi}^*}(\mathcal{G}_{\bm{\phi}^*}(\bm{H}_k + \Delta \bm{H}_k) + \Delta \bm{z}_k)$ being the reconstructed channel.


Finally the training and inference procedures of our proposed \ac{KD-EDN} for downlink beamforming is summarized in Algorithm 1. Note that this Algorithm produces a single channel-to-beamformer mapping under various SNR values, i.e., a mapping from the normalized estimated channel $\Tilde{\bm{H}}=[\Tilde{\bm{H}}_1,\dots,\Tilde{\bm{H}}_K]$ to the corresponding beamformer $\bm{W} = [\bm{W}_1,\dots,\bm{W}_K]$. To train such a mapping for different SNR values, we create a set $\mathcal{V}$ of noise variances corresponding to various SNRs. Then to obtain a normalized channel sample $\bm{H}^{(j)}$ in line 5 and line 11, for each user $k \in [K]$, we generate a channel sample $\Bar{\bm{H}}_k^{(j)}$ according to \eqref{mmwave_chan_user_k}, and randomly pick a noise variance $(\sigma_k^2)^{(j)}$ from $\mathcal{V}$, and form $\bm{H}_k^{(j)} = \frac{\Bar{\bm{H}}_k^{(j)}}{(\sigma_k^2)^{(j)}}$.

\setlength{\algorithmicindent}{2em}

\begin{algorithm} \label{alg_1}
\caption{\ac{KD-EDN} training and inference for downlink beamforming}
\begin{algorithmic} [1]
\STATE Randomly initialize three subnetwork parameters $\bm{\phi}^{(0)}$, $\bm{\theta}^{(0)}$ and $\bm{\psi}^{(0)}$, and let $\alpha=0$
\STATE \textbf{Training Stage:}   
    \STATE \quad \textbf{for} epoch $\ell=1,2,\ldots$ \textbf{do} \textit{// training of $(\bm{\phi}, \bm{\theta})$}
    \STATE \quad \quad Update the \ac{KD} weight $\alpha$
    \STATE \quad \quad Obtain training batch $\left\{(\bm{H}^{(j)},\Delta \bm{H}^{(j)}, \Delta \bm{z}^{(j)})\right\}_{j=1}^{N_{\text{b}}}$
    \STATE \quad \quad Compute the MMSE beamformer $\Tilde{\bm{W}}^{(j)}$ 
    \STATE \quad \quad Fix $\mathcal{F}_\theta$, update $\bm{\phi}$ according to \eqref{update_alter_en} using the \ac{KD} \\
    \quad \quad loss given by \eqref{proposed_kd_loss_func}-\eqref{mse_loss_goal}
    \STATE \quad \quad Fix $\mathcal{G}_\phi$, update $\bm{\theta}$ according to \eqref{update_alter_de} using the \ac{KD} \\
    \quad \quad loss given by \eqref{proposed_kd_loss_func}-\eqref{mse_loss_goal}
    \STATE \quad \textbf{end for}
    \STATE \quad \textbf{for} epoch $\ell=1,2,\ldots$ \textbf{do} \qquad \textit{// training of $\bm{\psi}$}
    \STATE \quad \quad Obtain training batch $\left\{(\bm{H}_k^{(j)},\Delta \bm{H}_k^{(j)}, \Delta \bm{z}_k^{(j)})\right\}_{j=1}^{N_{\text{b}}}$, \\
    \quad \quad for $k\in[K]$
    \STATE \quad \quad Update $\bm{\psi}$ using the loss given by \eqref{recovery_formu}
    \STATE \quad \textbf{end for}
    \STATE \quad \textbf{Output:} \ac{KD-EDN} parameters $(\bm{\phi}^*,\bm{\theta}^*,\bm{\psi^*})$
    
\STATE \textbf{Inference Stage:}
    \STATE \hspace{1em} \textbf{Each user} $k\in[K]$:
    \STATE \hspace{1em} - obtains the estimate of its normalized channel $\Tilde{\bm{H}}_k$
    \STATE \hspace{1em} - computes $\bm{z}_k=\mathcal{G}_{\bm{\phi}^*}(\Tilde{\bm{H}}_k)$ and send it to the \ac{BS}
    \STATE \hspace{1em} \textbf{The \ac{BS}}:
    \STATE \hspace{1em} - recovers latent variables $\hat{\bm{z}} = [\hat{\bm{z}}^T_1,\dots,\hat{\bm{z}}^T_K]^T$
    \STATE \hspace{1em} - computes the initial beamformer $\bm{W}_0=\mathcal{F}_{\bm{\theta}^*}(\hat{\bm{z}})$ and \\
    \hspace{1.65em} the recovered channels $\hat{\bm{H}}_k = \mathcal{J}_{\bm{\psi}^*}(\hat{\bm{z}}_k)$, $k \in [K]$
    \STATE \hspace{1em} - performs $Q_i$ steps of gradient ascent based on \eqref{gd_part_inference}
    \STATE \quad \textbf{Output:} Final beamforming solution $\bm{W}_{Q_i}$
\end{algorithmic}
\end{algorithm}

\vspace{-5pt}

\subsection{Simulation Results} \label{sr_perform}

\subsubsection{Simulation Setup} \label{simu_setup}
We consider a \ac{MISO} system comprising $K=16$ single-antenna users (i.e., $M=1$) and a \ac{BS} equipped with $N = 64$ transmit antennas, as well as a \ac{MIMO} system with $K = 4$ users, each equipped with $M = 4$ receive antennas. The channel samples are generated according to the sparse channel model in Sec.~\ref{sparse_chan}. Specifically, we set $L=10$ in \eqref{mmwave_chan_user_k} and $\sigma_{\gamma,k} = \sigma_{\beta,k}=0.1 \ \text{rad}$ in \eqref{multi_ang_calc}, $\forall k \in [K]$. All users have the same noise variance, i.e., $\sigma^2_k=\sigma^2$, $\forall k \in [K]$ in \eqref{receive_mimo_full_digit}. 
We set $\|\bm{W}\|_F^2=P=1$, and define $\text{SNR} \triangleq \frac{1}{\sigma^2}$.

The parameters of Algorithm 1 are set as follows. $\alpha$ in \eqref{proposed_kd_loss_func} is increased from 0 to 1 with a step-size $0.01$, and we set $N_b=128$ in Sec.~\ref{semi_amor_alter_train}, $\eta=10^{-4}$ in \eqref{update_alter}, $N_{\text{en}}=2$ in \eqref{update_alter_en}, $N_{\text{de}}=8$ in \eqref{update_alter_de}. Both the channel estimation error sample $\Delta \bm{H}$ and the feedback error sample $\Delta \bm{z}$ in \eqref{origin_loss_fun} are Gaussian distributed, such that $[\Delta \bm{H}]_{ij} \overset{\text{i.i.d.}}{\sim} \mathcal{CN}(0, \sigma^2_h)$, and $[\Delta \bm{z}]_{ij} \overset{\text{i.i.d.}}{\sim} \mathcal{CN}(0, \sigma^2_z)$, where $\sigma^2_h = \sigma^2_z=0.1$. 

Recall that our proposed \ac{KD-EDN} uses a combination of unsupervised and supervised loss functions, i.e., $L_{\alpha} = \alpha L^{\text{u}} + (1-\alpha) L^{\text{s}}$, where $\alpha$ gradually varies from 0 to 1. For comparison, we consider two baselines that employ Algorithm 1 with different loss functions: one termed ``Unsupervised'' uses the unsupervised loss function $L^{\text{u}}$ only, i.e., $\alpha=1$, and the other called ``Supervised'' uses the supervised loss function $L^{\text{s}}$ only, i.e., $\alpha=0$. The third baseline termed ``MMSE'' is a non-learning method. It first calculates the MMSE beamformer by using the estimated channel $\Tilde{\bm{H}}$, instead of the true channel $\bm{H}$. Then starting from $\bm{W}_0=\Tilde{\bm{W}}$, it performs $Q_i$ steps of gradient ascent as in \eqref{gd_part_inference} by using $\Tilde{\bm{H}}$. The final MMSE beamformer is $\bm{W}_{Q_i}$. Note that this baseline assumes perfect feedback, i.e., $\hat{\bm{H}} = \Tilde{\bm{H}}$. 



\subsubsection{Results} \label{fd_results}

\begin{figure*}
     \centering
     \justifying
     \begin{subfigure}[b]{0.48\textwidth}
         \centering
         \includegraphics[width=8.9cm]{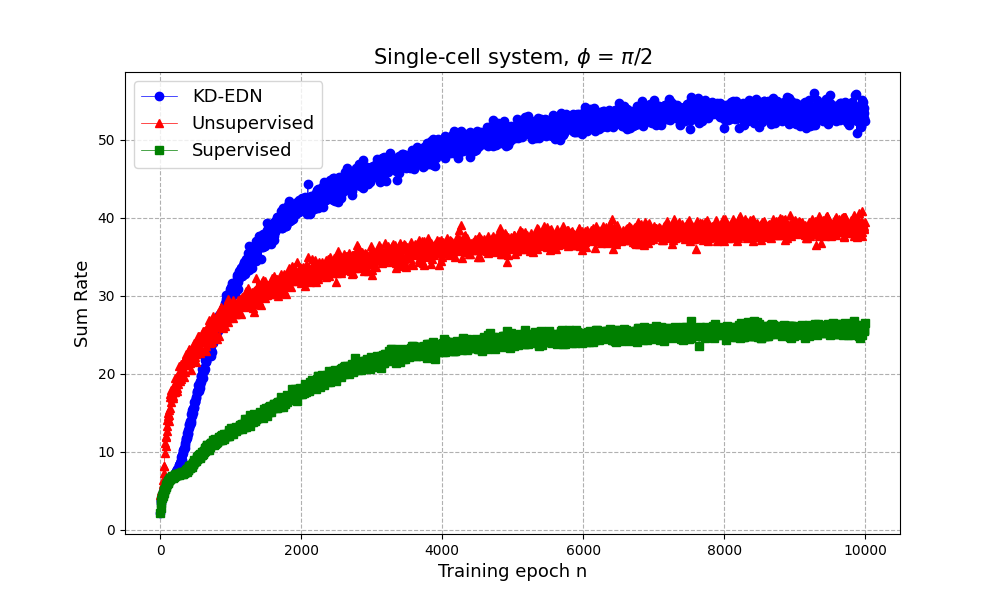}
            \caption{Single-cell system.}
		\label{sc_train}
     \end{subfigure}
     \begin{subfigure}[b]{0.48\textwidth}
        \centering
        \includegraphics[width=8.9cm]{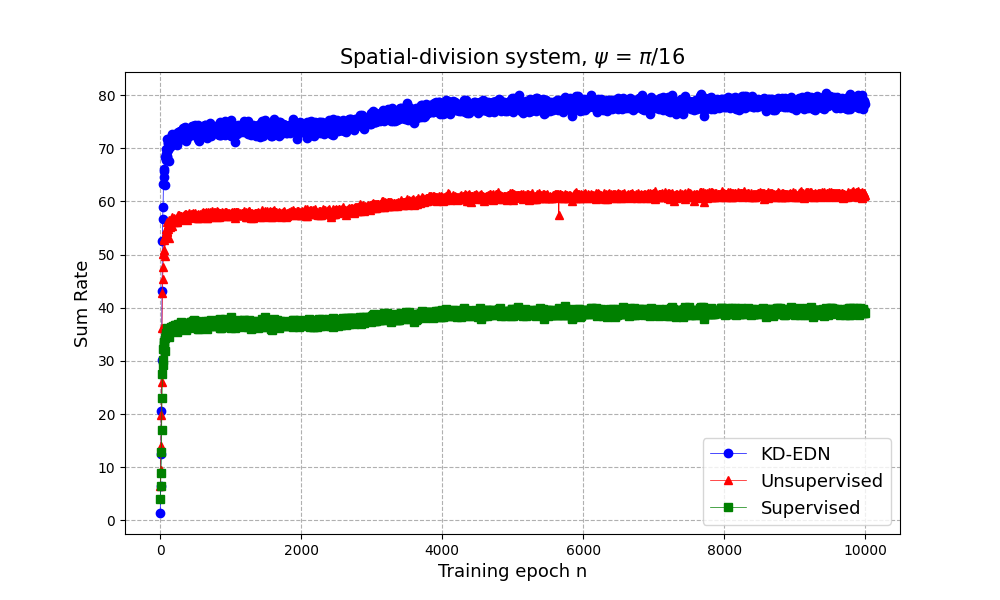}
            \caption{Spatial-division system.}
	    \label{sd_train}
    \end{subfigure}
        \caption{Convergence of joint training of encoder $\mathcal{G}_{\bm{\phi}}$ and  beamformer decoder $\mathcal{F}_{\bm{\theta}}$.}
        \label{train_result}
\end{figure*}

\begin{figure*}
     \centering
     \justifying
     \begin{subfigure}[b]{0.48\textwidth}
         \centering
         \raisebox{1mm}{
            \includegraphics[width=7.8cm]{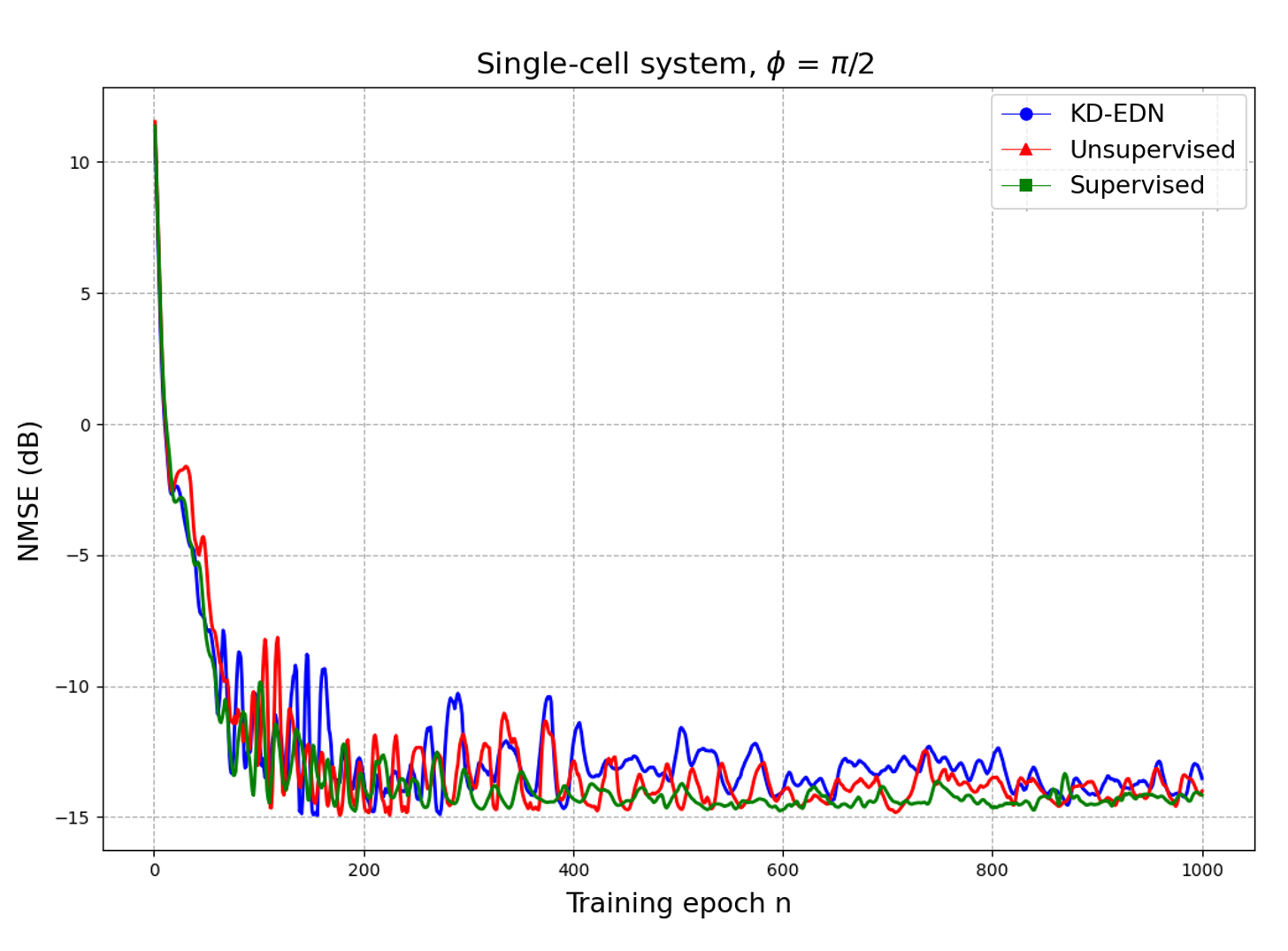}
         }
            \caption{Single-cell system.}
		\label{sc_chan_train}
     \end{subfigure}
     \begin{subfigure}[b]{0.48\textwidth}
        \centering
        \includegraphics[width=8.4cm]{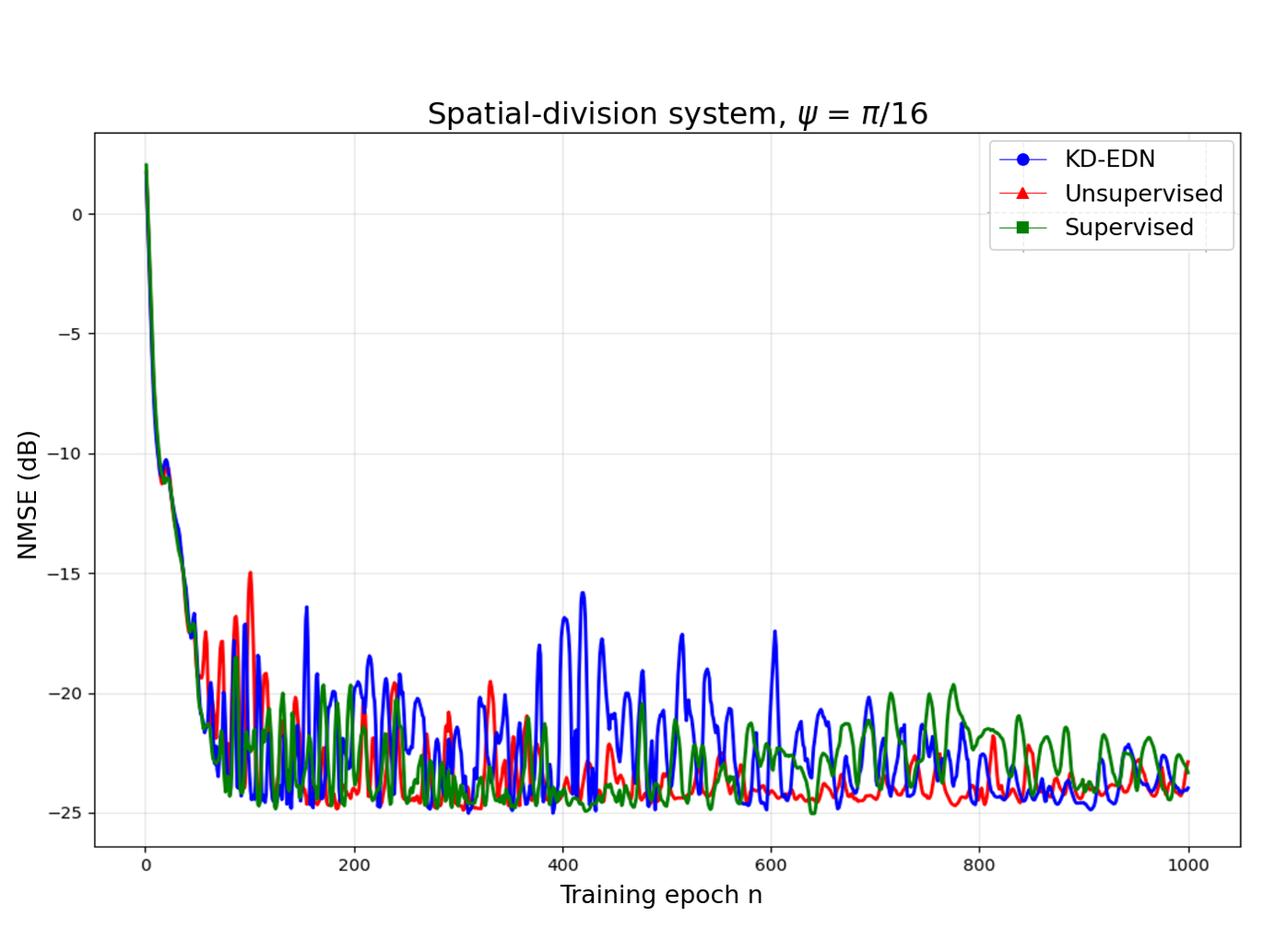}
            \caption{Spatial-division system.}
	    \label{sd_chan_train}
    \end{subfigure}
        \caption{Training convergence of channel decoder $\mathcal{J}_{\bm{\psi}}$.}
        \label{train_chan_result}
\end{figure*}

We first illustrate the training convergence of Algorithm 1 for both single-cell and spatial-division \ac{MISO} systems. The SNR is 15dB, and the number of gradient ascent steps in \eqref{gd_part_train} is $Q_t=5$. The sector angles for the single-cell and spatial-division systems are $\phi = \frac{\pi}{2}$ and $\psi = \frac{\pi}{16}$, respectively. Fig.~\ref{train_result} shows the convergence of the joint training of the encoder and the beamformer decoder $(\mathcal{G}_{\bm{\phi}},\mathcal{F}_{\bm{\theta}})$, where the sum rate versus training epoch is shown for the proposed \ac{KD-EDN} and two baselines. It is seen that the \ac{KD-EDN} outperforms both two baselines, demonstrating the benefits of KD. Furthermore, a comparison between Fig.~\ref{sc_train} and Fig.~\ref{sd_train} reveals that the spatial-division system achieves both higher sum-rate and faster convergence than the single-cell system. Intuitively, this is because users in a spatial-division system are better separated than those in a single-cell system, resulting in a better conditioned channel matrix $\bm{H}$.  
Fig.~\ref{train_chan_result} shows the training curves of the channel decoder $\mathcal{J}_{\bm{\psi}}$ for the three schemes, under both single-cell and spatial-division settings. It is seen that the two baseline schemes achieve similar channel recovery performance as the KD-EDN, and the spatial-division system has a much lower channel recovery error than the single-cell system.   

\begin{figure*}
     \centering
     \justifying
     \begin{subfigure}[b]{0.48\textwidth}
         \centering
         \includegraphics[width=8.3cm]{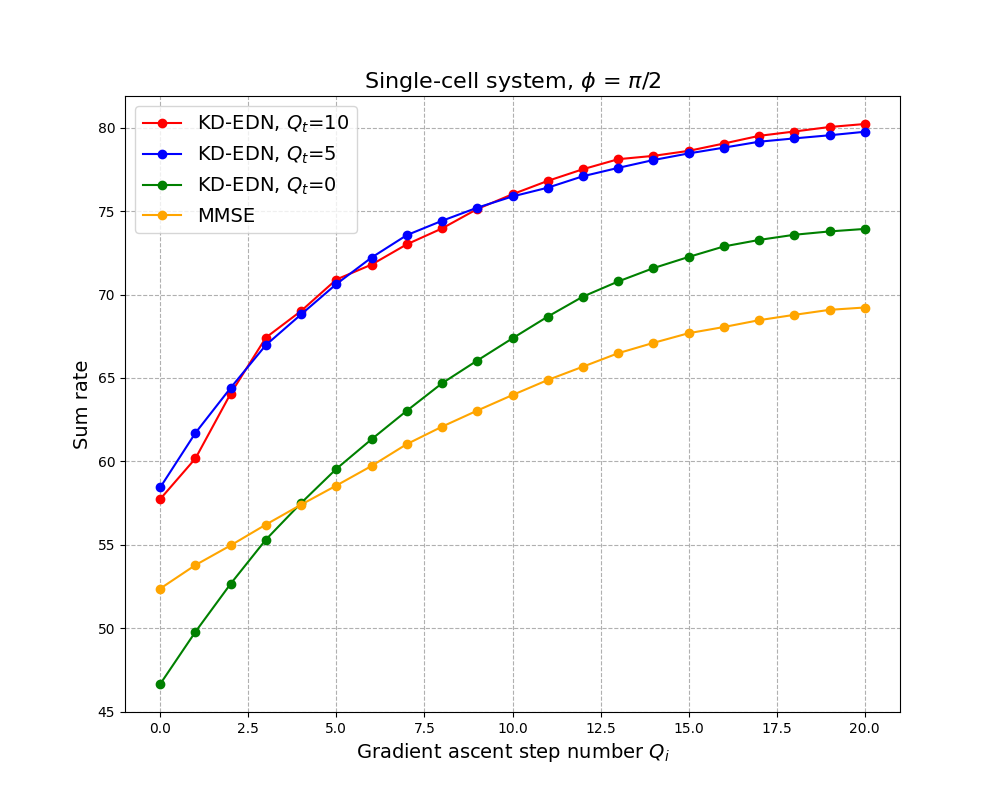}
            \caption{Single-cell system.}
		\label{sc_converge}
     \end{subfigure}
     \begin{subfigure}[b]{0.48\textwidth}
        \centering
        \includegraphics[width=8.3cm]{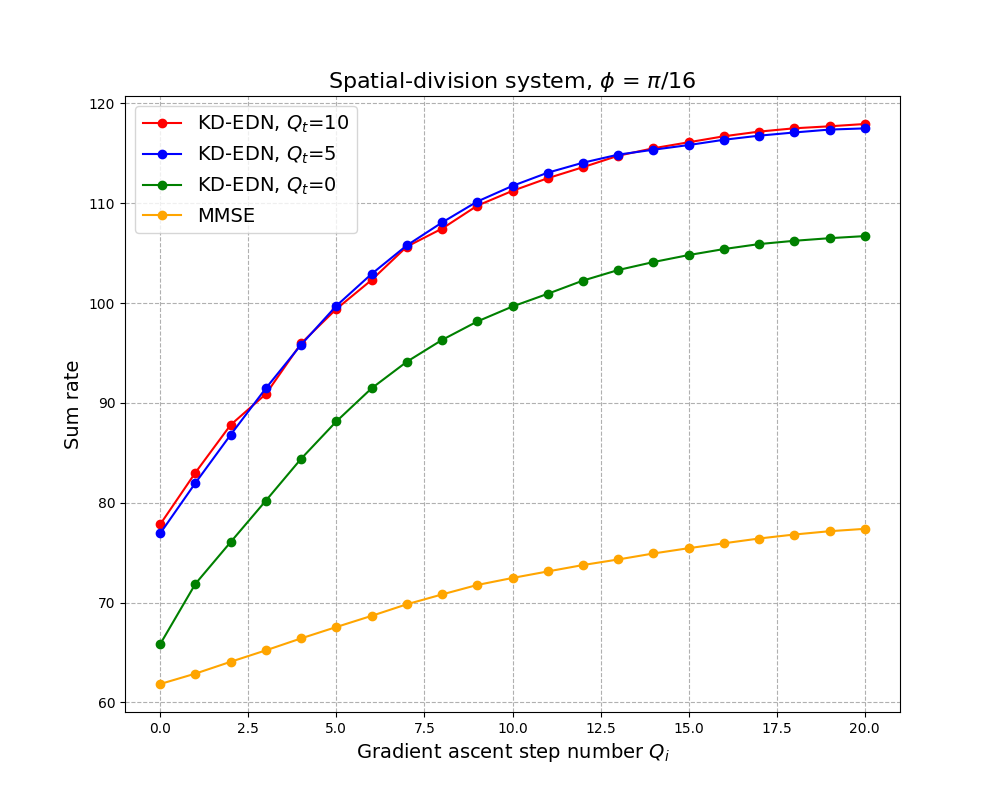}
            \caption{Spatial-division system.}
	    \label{sd_converge}
    \end{subfigure}
        \caption{Effects of the number of gradient ascent steps during training ($Q_t$) and inference ($Q_i$).}
        \label{converge_result}
\end{figure*}

Next, we evaluate how the numbers of gradient ascent steps, $Q_t$ during training as in \eqref{gd_part_train}, and $Q_i$ during inference as in \eqref{gd_part_inference}, affect the final sum-rate performance. The system setup is the same as that for Fig.~\ref{train_result} and Fig.~\ref{train_chan_result}. Three \ac{KD-EDN} models are trained using different numbers of gradient ascent steps in \eqref{gd_part_train}, i.e., $Q_t=0,5,10$, and then tested under varying number of $Q_i$ in \eqref{gd_part_inference}. Fig.~\ref{converge_result} shows the sum rate versus $Q_i$, using beamformers initialized by each \ac{KD-EDN} model. Also shown is the sum rate of the beamformer obtained by initializing \eqref{gd_part_inference} with the MMSE beamformer, with the true channel $\bm{H}_k$ replaced by its estimate $\Tilde{\bm{H}}_k$, and performs $Q_i$ gradient ascent steps in \eqref{gd_part_inference} with $\hat{\bm{H}}$ replaced by $\Tilde{\bm{H}}$. It is observed that the KD-EDNs with $Q_t=5$ and $Q_t=10$ significantly outperform the one with $Q_t=0$, demonstrating that the proposed semi-amortized training framework, i.e., integrating the analytical gradient ascent with neural networks to form the overall \ac{KD-EDN}, is effective in obtaining a high-quality mapping from the estimated channel $\Tilde{\bm{H}}$ to the beamformer $\bm{W}$ that yields high sum rate. Moreover, the gap between $Q_t=5$ and $Q_t=10$ is negligible, suggesting that during training a small number of gradient ascent steps is sufficient, reducing the corresponding additional training overhead, i.e., the computation of the intermediate gradient terms given in the Appendix. Moreover, during the inference stage, a larger $Q_i > Q_t$ can be employed in \eqref{gd_part_inference} to obtain a higher sum rate. Accordingly, we set $Q_t=5$ for training and $Q_i=10$ for inference in the subsequent simulations.

\begin{figure*}
     \centering
     \justifying
     \begin{subfigure}[b]{0.48\textwidth}
         \centering
         \includegraphics[width=8.0cm]{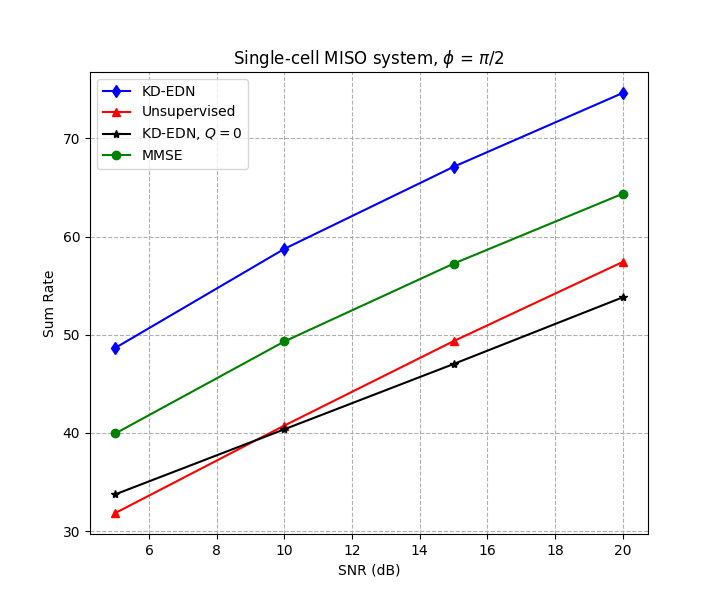}
            \caption{Single-cell MISO system.}
		\label{diff_snr_sc_miso}
     \end{subfigure}
     \begin{subfigure}[b]{0.48\textwidth}
        \centering
        \includegraphics[width=8.0cm]{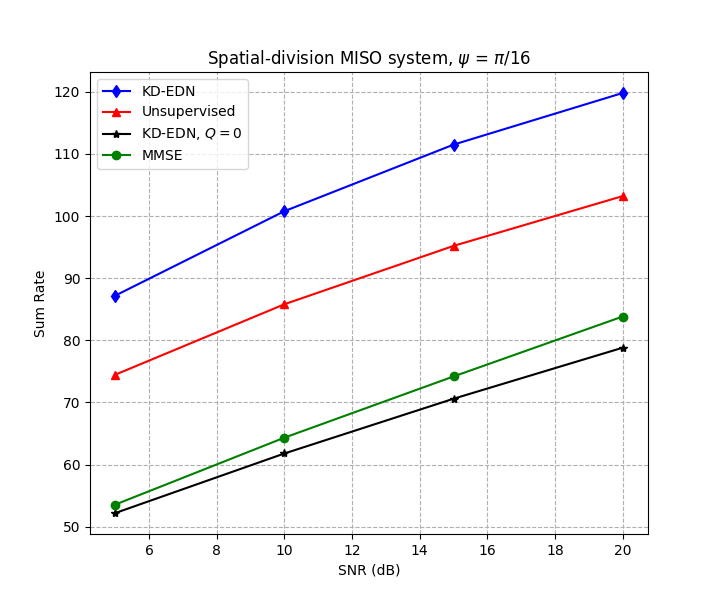}
            \caption{Spatial-division MISO system.}
	    \label{diff_snr_sd_miso}
    \end{subfigure}
        \caption{Sum-rate versus SNR in MISO systems.}
        \label{diff_snr_miso}
\end{figure*}
\begin{figure*}
     \centering
     \justifying
     \begin{subfigure}[b]{0.48\textwidth}
         \centering
         \includegraphics[width=8.0cm]{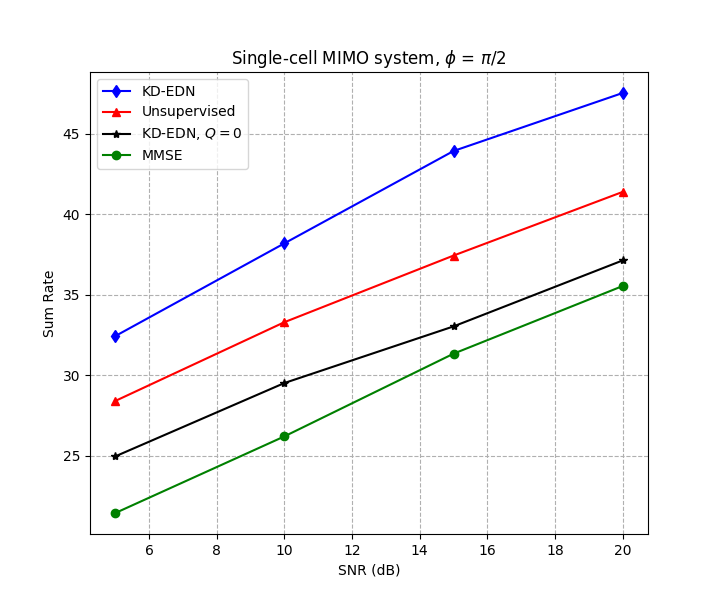}
            \caption{Single-cell MIMO system.}
		\label{diff_snr_sc_mimo}
     \end{subfigure}
     \begin{subfigure}[b]{0.48\textwidth}
        \centering
        \includegraphics[width=8.0cm]{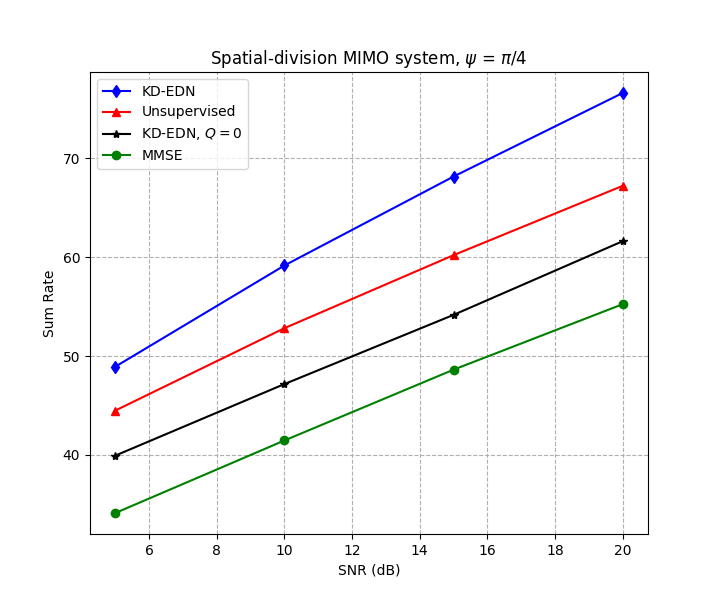}
            \caption{Spatial-division MIMO system.}
	    \label{diff_snr_sd_mimo}
    \end{subfigure}
        \caption{Sum-rate versus SNR in MIMO systems.}
        \label{diff_snr_mimo}
\end{figure*}

Finally, Fig.~\ref{diff_snr_miso} and Fig.~\ref{diff_snr_mimo} show the sum rate versus SNR for different schemes in \ac{MISO} and \ac{MIMO} systems, respectively. In addition to the ``Unsupervised'' and ``MMSE'' baselines, we also show the performance of a \ac{KD-EDN} with $Q_t=Q_i=0$, i.e., without semi-amortized learning. The sector angles are configured as: $\phi_{\text{MISO}} = \frac{\pi}{2},\phi_{\text{MIMO}} = \frac{\pi}{2}$ for single-cell systems, and $\psi_{\text{MISO}} = \frac{\pi}{16},\psi_{\text{MIMO}} = \frac{\pi}{4}$ for spatial-division systems. As discussed in Sec.~\ref{chan_dec_train}, a single \ac{KD-EDN} is trained for various SNR values by generating the normalized channel samples. In particular, $\mathcal{V}$ contains noise variance values corresponding to the SNR range 5-20dB, where $\text{SNR}=\frac{1}{\sigma^2}$. For each normalized channel sample $\bm{H}^{(j)}$, $\sigma_k^2=\sigma^2$, $k \in [K]$, and $\sigma^2$ is drawn uniformly from $\mathcal{V}$. The trained models are then evaluated across a range of SNR values. The results show a monotonic increase in sum rate with SNR, highlighting the generalization ability of the \ac{KD-EDN} framework across different SNR regimes. Moreover, both \ac{KD} and semi-amortized learning contribute to the superior performance of the proposed \ac{KD-EDN}, and the latter plays a more prominent role. For example, for the \ac{MISO} spatial-division case, \ac{KD} method alone (KD-EDN, $Q=0$) cannot outperform MMSE; whereas semi-amortized training alone (Unsupervised) can significantly outperform MMSE. For the \ac{MISO} single-cell case, neither of these two baselines can outperform MMSE; but when employed together, the \ac{KD-EDN} outperforms the MMSE. Note that since the MMSE baseline assumes that each estimated channel $\Tilde{\bm{H}}_k$ is perfectly transmitted from user $k$ to the BS, the corresponding curve represents an upper bound on its sum rate. A comparison between Fig.~\ref{diff_snr_miso} and Fig.~\ref{diff_snr_mimo} further illustrates that \ac{MISO} systems achieve higher sum rates than the \ac{MIMO} counterparts, owing to a higher multi-user diversity gain and a greater scheduling flexibility. 

\vspace{-3pt}
\section{Applications to Hybrid Beamforming} \label{extend_to_hybrid}

In this section, we apply the \ac{KD-EDN} beamforming scheme developed in Sec.~\ref{DL_framework} to hybrid beamforming scenarios, in both far-field and near-field channels. 

\vspace{-3pt}
\subsection{Hybrid Beamforming in Far-field Channels} \label{hybrid}
Conventional fully-digital beamforming architectures require a dedicated \ac{RF} chain for each antenna element, leading to high cost and high power consumption. The hybrid beamforming solution consisting of a low-dimensional digital beamformer and an analog phase-shifter-based beamformer effectively reduces the number of \ac{RF} chains and enhances the energy efficiency \cite{alkhateeb2014channel,sohrabi2016hybrid}. 

Consider a downlink \ac{MIMO} hybrid beamforming system, where a \ac{BS} equipped with $N$ transmit antennas and $N_{\text{RF}}$ \ac{RF} chains serves $K$ users. Each user is equipped with $M$ antennas. Typically, in hybrid beamforming it is assumed that $KM \leq N_{\text{RF}} \ll N$ \cite{sohrabi2016hybrid}. The received signal at user $k$ is still given by \eqref{receive_mimo_full_digit}, with the beamformer matrix $\bm{W}_k \in \mathbb{C}^{N \times M}$ of each user $k$ composed of two components --- an analog component $\bm{W}^a \in \mathbb{C}^{N \times N_{\text{RF}}}$ that is common to all users, and a user-specific digital component $\bm{W}_k^D \in \mathbb{C}^{N_{\text{RF}} \times M}$, i.e., $\bm{W}_k = \bm{W}^a \bm{W}_k^D$. 


In particular, each column of the analog beamformer $\bm{W}^a$ is a beam-steering vector parameterized by a steering angle. Let $\bm{\beta} \triangleq \{\beta_1,\dots,\beta_{N_{\text{RF}}}\}$ denote $N_{\text{RF}}$ steering angles, and $d$ and $\lambda$ denote the antenna spacing and carrier wavelength, respectively. The analog beamformer is given by 
\begin{align} \label{steer_vector_hybrid}
\bm{W}^a (\bm{\beta}) = \left[\bm{a}_N \left(\beta_1, d / \lambda \right),\dots,\bm{a}_N \left(\beta_{N_{\text{RF}}}, d / \lambda \right)\right]. 
\end{align}
We now specify the steering angles $\bm{\beta}$ for two scenarios of user geographical distribution in Fig.~\ref{sys_graph}. In single-cell case with an angular coverage of $[-\frac{\phi}{2}, \frac{\phi}{2}]$, the steering angles are set as $\beta_n = -\frac{\phi}{2}+\frac{(n-\frac{1}{2})\phi}{N_{\text{RF}}}$, for $n \in [N_{\text{RF}}]$, so that the resulting beam directions are evenly spread across the sector. In spatial-division case where each user $k$ is located within the angular interval $[\eta_K(k)-\frac{\psi}{2},\eta_K(k)+\frac{\psi}{2}]$, $\eta_K(k) = -\frac{\pi}{2} + \frac{(k-\frac{1}{2})\pi}{K}$, for simplicity we assume that $N_{\text{RF}}=KM$ and set the steering angles as $\beta_{(k-1)M+m} = \eta_K(k) -\frac{\psi}{2}+\frac{(m-\frac{1}{2})\psi}{M}$, where $k\in[K], m\in[M]$, so that the beam directions are uniformly distributed within each user's designated sector.


Since the analog beamformer $\bm{W}^a(\bm{\beta})$ is pre-determined, it can be absorbed into the channel matrix $\Bar{\bm{H}}_k$ in \eqref{receive_mimo_full_digit}, resulting in the following effective channel model
\begin{align} \label{mimo_far_hy_sig_model}
\bm{y}_k = \Bar{\bm{G}}_k \bm{W}^D_k \bm{x}_k + \Bar{\bm{G}}_k \sum_{i \neq k} \bm{W}^D_i \bm{x}_i + \bm{n}_k,
\end{align}
where $\Bar{\bm{G}}_k \triangleq \Bar{\bm{H}}_k \bm{W}^a(\bm{\beta}) \in \mathbb{C}^{M \times N_{\text{RF}}}$, $k \in [K]$. Define the normalized effective channel matrix of user $k$ as $\bm{G}_k \triangleq \frac{\Bar{\bm{G}}_k}{\sigma_k^2}$. Further, denote the overall effective channel matrix by $\bm{G}=[\bm{G}_1,\dots,\bm{G}_K]$ and the overall digital beamforming matrix by $\bm{W}^D=[\bm{W}^D_1,\dots,\bm{W}^D_K] \in \mathbb{C}^{N_{\text{RF}} \times KM}$. The achievable sum rate of the system is then given by
\begin{align} \label{mimo_sr_hybrid}
R_\text{sum}(\bm{G},\bm{W}^D)=\sum_{k=1}^K  \log_2 \lvert \bm{I}_M + \bm{V}_k^{-1} \bm{G}_k \bm{W}^D_k (\bm{W}^D_k)^H \bm{G}_k^H \rvert, 
\end{align}
where $\bm{V}_k = \bm{I}_M + \bm{G}_k \left( \sum_{i \neq k} \bm{W}^D_i (\bm{W}^D_i)^H \right) \bm{G}_k^H$ is the covariance matrix of interference-plus-noise term for user $k$. 

For the special case of a \ac{MISO} system, i.e., $M=1$, the channel between the \ac{BS} and user $k$ is denoted by $\Bar{\bm{h}}^T_k \in \mathbb{C}^{1 \times N}$. Given the analog beamformer $\bm{W}^a(\bm{\beta}) \in \mathbb{C}^{N \times N_{\text{RF}}}$, the effective channel vector is $(\Bar{\bm{g}}_k)^T = (\Bar{\bm{h}}_k)^T \bm{W}^a(\bm{\beta}) \in \mathbb{C}^{1 \times N_{\text{RF}}}$. Assume that a digital beamformer $\bm{w}^D_k \in \mathbb{C}^{N_{\text{RF}}}$ is applied to transmit the data symbol $x_k$. The received signal at user $k$ is
\begin{align} \label{receive_miso_hybrid}
y_k = (\Bar{\bm{g}}_k)^H \sum_{i=1}^{K} \bm{w}^D_i x_i + n_k, \ k \in [K],
\end{align}
where $n_k \sim \mathcal{CN}(0,\sigma_k^2)$. Hence the normalized effective channel of user $k$ is given by $\bm{g}^T_k=\frac{\Bar{\bm{g}}^T_k}{\sigma_k^2}$. Denote $\bm{G}=[\bm{g}_1,\dots,\bm{g}_K]$, and $\bm{W}^D=[\bm{w}^D_1,\dots,\bm{w}^D_K]$. The achievable sum rate is then
\begin{align} \label{miso_sr_hybrid}
R_\text{sum}(\bm{G},\bm{W}^D) = \sum_{k=1}^K \log_2\left(1+\frac{\left|\bm{g}_k^H \bm{w}^D_k \right|^2}{1+\sum_{i \neq k} \left|\bm{g}_k^H \bm{w}^D_i \right|^2}\right). 
\end{align}

During the pilot phase, using the analog beamformer $\bm{W}^a$, the \ac{BS} transmits pilot symbols to all users. Each user $k$ estimates its effective channel $\bm{G}_k$, and feeds the estimate $\Tilde{\bm{G}}_k$ to the encoder $\mathcal{G}_{\bm{\phi}}$ to produce the latent variable $\bm{z}_k$, which is then compressed and sent back to the \ac{BS}. The \ac{BS} first recovers the latent variables $\hat{\bm{z}}=[\hat{\bm{z}}_1,\dots,\hat{\bm{z}}_K]$. Then using the channel decoder $\mathcal{J}_{\bm{\psi}}$, it computes the effective channels $\hat{\bm{H}}_k=\mathcal{J}_{\bm{\psi}}(\hat{\bm{z}}_k)$. Finally it computes the digital beamformer by initializing with the beamformer decoder $\bm{W}^D_0 = \mathcal{F}_{\bm{\theta}}(\hat{\bm{z}})$, and performing $Q_i$ steps of gradient ascent based on \eqref{gd_part_inference} using the recovered channel $\hat{\bm{H}} = [\hat{\bm{H}}_1,\dots,\hat{\bm{H}}_K]$, to obtain $\bm{W}^D_{Q_i}$ which is employed during the data transmission phase. 


Given the pre-determined steering angles $\bm{\beta}$, we aim to jointly learn the mappings $g: \Tilde{\bm{G}}_k \rightarrow \bm{z}_k$ and $f: \hat{\bm{z}} \rightarrow \bm{W}^D$, such that the average sum rate is maximized, i.e.,
\begin{align} \label{unified_formulation_hybrid}
&\max_{\substack{\bm{z}_k = g(\Tilde{\bm{G}}_k), \\ \bm{W}^D= f(\hat{\bm{z}})}} \ \mathbb{E}_{\bm{G},\Delta \bm{G},\Delta \bm{z}} \left[R_{\text{sum}}(\bm{G},\bm{W}^D)\right], \notag \\
& \qquad \qquad \text{s.t.} \ \ \| \bm{W}^a(\bm{\beta}) \cdot \bm{W}^D \|_{\text{F}}^2 \leq P,
\end{align}
where $\Tilde{\bm{G}}_k$ denotes the estimated effective channel at user $k$. Hence for a given user geographical distribution scenario, the analog beamformer $\bm{W}^a(\bm{\beta})$ can be computed by \eqref{steer_vector_hybrid}. Using the training batch $\{(\bm{G}^{(j)}, \Delta \bm{G}^{(j)}, \Delta \bm{z}^{(j)})\}_{j=1}^{N_b}$, where $\bm{G}^{(j)}=\bm{H}^{(j)} \bm{W}^a(\bm{\beta})$, we can then apply Algorithm 1 to train a \ac{KD-EDN} and use it to compute the digital beamformer $\bm{W}^D$ for downlink hybrid beamforming. Note that in this case, the normalization layer operation should be modified as follows: Taking the \ac{MISO} system as an example, it first reshapes the output vector of the beamformer decoder, denoted by $\Tilde{\bm{w}}^D$, into a complex-valued matrix $\Tilde{\bm{W}}^D = \text{vec}_w^{-1} \left(\Tilde{\bm{w}}^D \left[1 : K N_{\text{RF}} \right]\right) + j \cdot \text{vec}_w^{-1} \left(\Tilde{\bm{w}}^D \left[K N_{\text{RF}}+1 : 2 K N_{\text{RF}} \right]\right)$, where $\text{vec}_w^{-1}$ denotes the inverse vectorization operation that reshapes a real-valued vector of size $K N_{\text{RF}}$ into a matrix of size $N_{\text{RF}} \times K$. The resulting $\Tilde{\bm{W}}^D$ is then normalized to satisfy the overall power constraint of the  hybrid beamformer, i.e., 
\begin{align} \label{norm_digi_bf_ff}
\bm{W}^D = \sqrt{P} \cdot \frac{\Tilde{\bm{W}}^D}{\|\bm{W}^a(\bm{\beta}) \cdot \Tilde{\bm{W}}^D \|_F}.
\end{align}

\subsection{Hybrid Beamforming in Near-field Channels} \label{near_field}

With the deployment of larger antenna apertures and/or wider communication bandwidths, the beam splitting phenomenon emerges due to the spherical wavefront nature of near-field propagation \cite{wang2024beamfocusing, cui2024near, elbir2023near}. Specifically, beams at different frequencies focus on distinct locations, causing a severe loss of the beamforming gain. 

In the near-field, since the wavefronts are spherical, the phase varies with both distance and angle; whereas in the far-field, the wavefronts are planar and the phase varies with only the angle. For the near-field channel model, we consider the piecewise-far-field model in \cite{cui2024near}, where the \ac{BS} antenna array is partitioned into multiple sub-arrays, and it is assumed that a user lies within the far-field region of each sub-array. 

In near-field systems which operate in the \ac{THz} bands, severe reflection losses result in significantly higher attenuation for \ac{NLoS} paths compared to \ac{LoS} paths \cite{wang2024beamfocusing}. Consequently, the channel between a transmit antenna and a receive antenna consists of only a single dominant \ac{LoS} path; and according to the free-space Maxwell equation \cite{zhou2015spherical}, it is given by $\Bar{h} = \alpha e^{-j \frac{2\pi }{\lambda} r}$, where $\lambda$ and $r$ denote the wavelength and the distance between the transmit and receive antennas, respectively, and $\alpha \sim \mathcal{CN}(0,1)$ denotes the complex path loss. For simplicity, we consider the \ac{MISO} case, where a \ac{BS} equipped with $N$ transmit antennas and $N_{\text{RF}}$ RF chains serves $K$ single-antenna users, where $N_{\text{RF}} = K$. In particular, the $N$ transmit antennas are partitioned into $S$ sub-arrays, each consisting of $n$ antennas, such that $N=Sn$. Let $r^{s,q}_{k}$ denote the distance from the $((s-1)n+q)^{\text{th}}$ transmit antenna to the $k^{\text{th}}$ user, $s \in [S]$, $q \in [n]$. The sub-channel vector between the $s^{\text{th}}$ \ac{BS} sub-array and the $k^{\text{th}}$ user is then given by
\begin{align}  \label{nf_chan_subarray}
(\Bar{\bm{h}}^s_{k})^T = \alpha_k \left[e^{-j \frac{2\pi }{\lambda} r^{s,1}_k}, \dots, e^{-j \frac{2\pi }{\lambda} r^{s,n}_k}\right] \in \mathbb{C}^{1 \times n},
\end{align}
where $s \in [S]$. Let $r_k^s$ and $\theta_k^s$ denote the distance and angle from the first antenna (i.e., $q=1$) of the $s^{\text{th}}$ \ac{BS} sub-array to the $k^{\text{th}}$ user, respectively. According to the first-order approximation of the distance parameter in \cite{cui2024near}, we have
\begin{align} \label{one_approx_chan}
r^{s,q}_{k} \approx r_k^s + (q-1) d \sin{\theta_k^s}, \ s\in[S], \ q\in[n],
\end{align}
where $d$ denotes the antenna array spacing. Substituting \eqref{one_approx_chan} into \eqref{nf_chan_subarray}, we have
\begin{align}  \label{nf_chan_subarray_approx}
(\Bar{\bm{h}}^s_{k})^T = \sqrt{n} \alpha_k \cdot e^{-j \frac{2\pi}{\lambda} r^s_{k}} \cdot \bm{a}^H_n \left(\theta^s_{k}, d / \lambda \right) \in \mathbb{C}^{1 \times n},
\end{align}
where $s \in [S]$, and the array response vector $\bm{a}_n\left(\theta,\frac{d}{\lambda}\right)$ is defined in \eqref{array_response_vec}. Compared to the model in \eqref{mmwave_chan_user_k}, the near-field channel model in \eqref{nf_chan_subarray_approx} incorporates a distance-dependent phase term $e^{-j \frac{2\pi}{\lambda} r^s_{k}}$ for each sub-array $s$. The reason is that far-field channel models assume equal distances from all transmit antennas to a receiving terminal; whereas for near-field channel models, different transmit sub-arrays have different distances to the same terminal. As a result, the total phase in \eqref{nf_chan_subarray_approx} is composed of two components: the first one depends on the wavelength $\lambda$, or equivalently, the frequency $f_c$ and the distance $r^s_{k}$, and the second one depends on the angle $\theta^s_{k}$ and is independent of the frequency. 


The analog beamformer matrix is denoted by $\bm{W}^a = [\bm{v}_1,\dots,\bm{v}_K] \in \mathbb{C}^{N \times K}$ (recall that $N_{\text{RF}}=K$), where each analog beamformer $\bm{v}_k$ is also partitioned into $S$ sub-vectors. Corresponding to \eqref{nf_chan_subarray_approx}, the $s^{\text{th}}$ sub-vector $\bm{v}^s_k$ consists of two components: one produced by a frequency-dependent \ac{TTD} module to compensate for the phase term $e^{-j \frac{2\pi}{\lambda} r^s_{k}}$ in \eqref{nf_chan_subarray_approx}, and another produced by a frequency-independent phase shifter (PS) module to compensate for the array response term $\bm{a}^{H}_n \left(\theta^s_{k},\frac{d}{\lambda}\right)$ in \eqref{nf_chan_subarray_approx}. That is,
\begin{align} \label{nf_bf_subarray_miso}
\bm{v}^{s}_k = \sqrt{n /N} \cdot e^{j \frac{2\pi}{\lambda} \mu^s_{k}} \cdot \bm{a}_n \left(\eta^s_{k},d / \lambda \right) \in \mathbb{C}^{n \times 1}, s\in[S],
\end{align}
where $\mu^s_{k} \geq 0$ denotes the adjustable delay parameter of the $s^{\text{th}}$ \ac{TTD} element connected to the $k^{\text{th}}$ RF chain, and $\eta^s_{k}$ denotes the corresponding phase-shift parameter of \ac{PS} elements connected to the same RF chain. The $k^{\text{th}}$ analog beamformer is then given by $\bm{v}_k = [(\bm{v}^{1}_k)^T,\dots,(\bm{v}^{S}_k)^T]^T \in \mathbb{C}^{N \times 1}$, which is associated with the $k^{\text{th}}$ RF chain and beam-focusing towards the $k^{\text{th}}$ user. Specifically, denote by $\Bar{\bm{h}}^T_k = [(\Bar{\bm{h}}^1_k)^T,\dots,(\Bar{\bm{h}}^S_k)^T] \in \mathbb{C}^{1 \times N}$ the channel between the \ac{BS} and user $k$. Then the normalized beamforming gain of the analog beamformer $\bm{v}_k$ on the channel $\Bar{\bm{h}}^T_k$ is given by
\begin{align} \label{norma_beam_gain}
b_{k} &= \frac{1}{\sqrt{N |\alpha_{k}| }} |\Bar{\bm{h}}^H_{k} \bm{v}_{k}| = \frac{1}{\sqrt{N|\alpha_{k}|}} \left|\sum_{s=1}^{S} (\Bar{\bm{h}}^s_{k})^H \bm{v}^{s}_{k}\right| \notag \\
&= \frac{1}{N} \left| \sum_{s=1}^{S} e^{j \frac{2\pi }{\lambda} (\mu^s_{k}+r^s_{k})} \sum_{q=1}^{n} e^{j \pi (q-1) (\sin{\eta^s_{k}}+\sin{\theta^s_{k}})} \right|.
\end{align}
Since $\mu^s_{k} \geq 0$, $b_{k}$ is maximized when the parameters of $\bm{v}^{s}_k$ are chosen as \cite{cui2024near}
\begin{align} \label{opt_eta_mu}
\eta^s_{k} = - \theta^s_{k}, \ \mu^s_{k} = \max_{s\in[S]}\{r^s_{k}\} - r^s_{k}, \ s\in[S].
\end{align}
Substituting \eqref{opt_eta_mu} into \eqref{nf_bf_subarray_miso} yields the optimal analog beamformer, denoted by $\bm{v}_{k}(\bm{p}_k) \in \mathbb{C}^{N \times 1}$, $\bm{p}_k = \{(r_k^s, \theta_k^s), s \in [S]\}$. The overall analog beamformer is then given by
\begin{align} \label{overall_ana_bf_nf}
\bm{W}^a(\bm{P}) = [\bm{v}_1(\bm{p}_1),\dots,\bm{v}_K(\bm{p}_K)] \in \mathbb{C}^{N \times K}.
\end{align}
Define the effective channel vector for user $k$ as $\Bar{\bm{g}}^T_k \triangleq \Bar{\bm{h}}^T_k \bm{W}^a(\bm{P}) \in \mathbb{C}^{1 \times N_{\text{RF}}}$, and assume that a digital beamformer $\bm{w}^D_k \in \mathbb{C}^{N_{\text{RF}} \times 1}$ is applied to transmit the data symbol $x_k$ intended for user $k$. Then the received signal at user $k$ is 
\begin{align} \label{receive_miso_nr_hybrid}
y_k = (\Bar{\bm{g}}_k)^H \sum_{i=1}^{K} \bm{w}^D_i x_i + n_k, \ k \in [K],
\end{align}
where $n_k \sim \mathcal{CN}(0,\sigma_k^2)$. Hence the normalized effective channel of user $k$ is given by $\bm{g}^T_k=\frac{\Bar{\bm{g}}^T_k}{\sigma_k^2}$. Denote $\bm{G}=[\bm{g}_1,\dots,\bm{g}_K]^T \in \mathbb{C}^{K \times N_{\text{RF}}}$, and $\bm{W}^D=[\bm{w}^D_1,\dots,\bm{w}^D_K] \in \mathbb{C}^{N_{\text{RF}} \times K}$. The achievable sum rate is then given by
\begin{align} \label{miso_sr_nr_hybrid}
R_\text{sum}(\bm{G},\bm{W}^D) = \sum_{k=1}^K \log_2\left(1+\frac{\left|\bm{g}_k^H \bm{w}^D_k \right|^2}{1+\sum_{i \neq k} \left|\bm{g}_k^H \bm{w}^D_i \right|^2}\right). 
\end{align}
Note that the choice of the analog beamformer $\bm{W}^a(\bm{P})$ in \eqref{norma_beam_gain} maximizes the magnitude of the diagonal elements of the square matrix $\bm{G}$, i.e., the designated signal power at each user. We then design the digital beamformer $\bm{W}^D$ to maximize the sum rate by mitigating the inter-user interference.

Similar to the far-field case in Sec.~\ref{hybrid}, we aim to jointly learn the mappings $g: \Tilde{\bm{g}}_k \rightarrow \bm{z}_k$ and $f: \hat{\bm{z}} \rightarrow \bm{W}^D$, such that the average sum rate is maximized, i.e.,
\begin{align} \label{unified_formulation_nf}
&\max_{\substack{\bm{z}_k = g(\Tilde{\bm{g}}_k), \\ \bm{W}^D= f(\hat{\bm{z}})}} \ \mathbb{E}_{\bm{G},\Delta \bm{G},\Delta \bm{z}} \left[R_{\text{sum}}(\bm{G},\bm{W}^D)\right], \notag \\ 
& \qquad \qquad \text{s.t.} \ \ \| \bm{W}^a(\bm{P}) \cdot \bm{W}^D \|_{\text{F}}^2 \leq P,
\end{align}
where $\Tilde{\bm{g}}_k$ denotes the estimated effective channel for user $k$. Using the training batch $\{(\bm{G}^{(j)}, \Delta \bm{G}^{(j)}, \Delta \bm{z}^{(j)})\}_{j=1}^{N_b}$, where $\bm{G}^{(j)}=\bm{H}^{(j)} \bm{W}^{a}(\bm{P}^{(j)})$, we can employ Algorithm 1 to train a \ac{KD-EDN} and use it to compute the digital beamformer $\bm{W}^D$ for downlink near-field hybrid beamforming, with the normalization layer operation given by \eqref{norm_digi_bf_ff}, where $\bm{W}^a(\bm{\beta})$ is replaced by $\bm{W}^a(\bm{P})$. Note that for the far-field case, $\bm{W}^a(\bm{\beta})$ is a pre-determined constant matrix, whereas for the near-field case, $\bm{W}^a(\bm{P})$ is a function of the input sample.


\vspace{-3pt}
\subsection{Simulation Results} \label{sr_perform_hybrid}

\subsubsection{Simulation Setup} \label{simu_setup_hybrid}
Consider a \ac{MISO} system with $K=16$ single-antenna users (i.e., $M=1$) and a \ac{BS} equipped with $N = 64$ transmit antennas and $N_{\text{RF}}=16$ RF chains. In far-field cases, channel samples are generated according to Sec.~\ref{sparse_chan}, using the same parameters as in Sec.~\ref{simu_setup}, while the pre-determined analog beamformer is constructed following \eqref{steer_vector_hybrid}. In near-field cases, both the channel and analog beamformer samples are generated according to \eqref{nf_chan_subarray_approx} and \eqref{nf_bf_subarray_miso}, respectively, with $\lambda=3 \times 10^{-3}$m, $S=16$ and $n=4$ in \eqref{nf_chan_subarray_approx}. The Rayleigh distance is $R=\frac{N^2 \lambda}{2}=6.144\text{m}$, within which the near-field region is typically defined. We assume that each user's distance from the \ac{BS} is Gaussian distributed, such that $r_k \sim \mathcal{N}(r_c,\sigma_r^2)$, $\forall k\in[K]$, where $r_c = 3\text{m}$ and $\sigma_r^2=1$. Each user's angle $\theta_k$ is generated in the same manner as in the far-field case. Based on $(r_k,\theta_k)$, the subarray-wise user locations $\bm{p}_k = \{(r_k^s, \theta_k^s), s \in [S]\}$ are then computed as
\begin{align} \label{subarray_para_calc}
&r_k^s = \sqrt{r_k^2 + ((s-1)nd)^2 - 2 (s-1)nd r_k \sin{\theta_k}}, \notag \\
&\sin{\theta_k^s} = \frac{r_k \sin{\theta_k} - (s-1)nd}{r_k^s},
\end{align}
where $d=\frac{\lambda}{2}$. All other parameters remain identical to those in Sec.~\ref{simu_setup}.

To benchmark the proposed \ac{KD-EDN} scheme, as in Sec.~\ref{fd_results}, we consider three baselines: the ``Unsupervised'' one uses the unsupervised loss function $L^\text{u}$ with $Q_t=5,Q_i=10$; another termed ``KD-EDN, $Q=0$'' adopts a trained \ac{KD-EDN} with $Q_t=Q_i=0$; the final termed ``MMSE'' computes the MMSE digital beamformer using the estimated effective channel $\Tilde{\bm{G}}$, followed by $Q_i=10$ steps of gradient ascent based on $\Tilde{\bm{G}}$ to obtain the final beamformer $\bm{W}^D_{Q_i}$.


\subsubsection{Results} \label{result_hybrid}
\begin{figure*}
     \centering
     \justifying
     \begin{subfigure}[b]{0.48\textwidth}
         \centering
         \includegraphics[width=8.2cm]{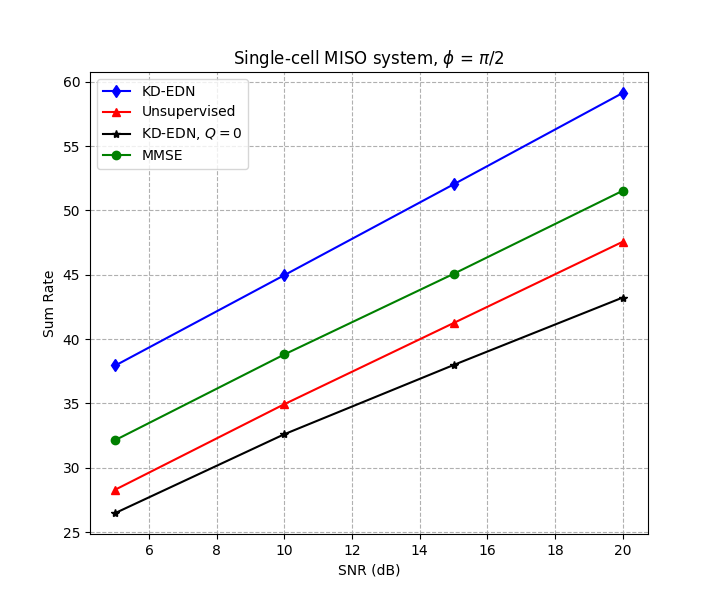}
            \caption{Single-cell MISO system.}
		\label{diff_snr_sc_miso_ff}
     \end{subfigure}
     \begin{subfigure}[b]{0.48\textwidth}
        \centering
        \includegraphics[width=8.2cm]{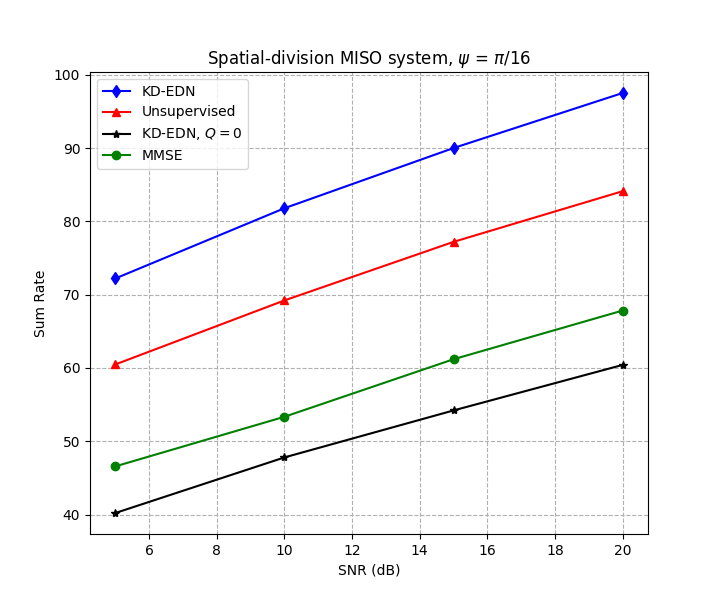}
            \caption{Spatial-division MISO system.}
	    \label{diff_snr_sd_miso_ff}
    \end{subfigure}
        \caption{Sum-rate versus SNR in far-field hybrid beamforming systems.}
        \label{diff_snr_ff}
\end{figure*}

\begin{figure*}
     \centering
     \justifying
     \begin{subfigure}[b]{0.48\textwidth}
         \centering
         \includegraphics[width=8.2cm]{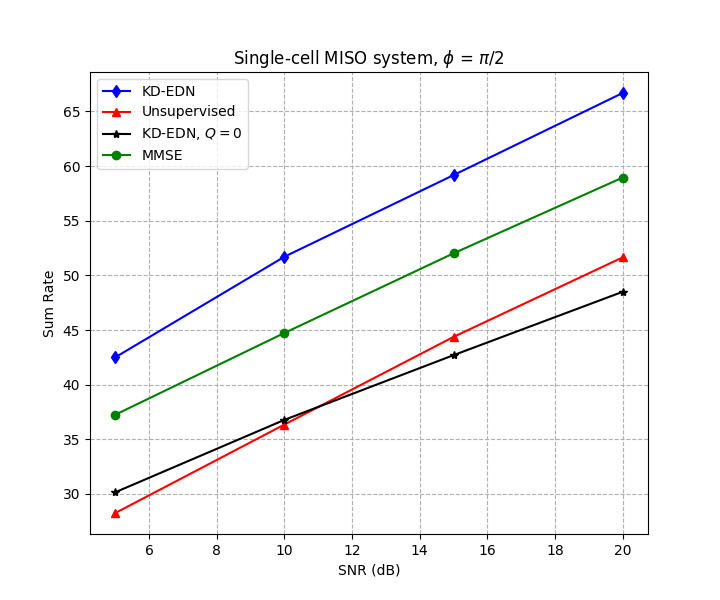}
            \caption{Single-cell MISO system.}
		\label{diff_snr_sc_miso_nf}
     \end{subfigure}
     \begin{subfigure}[b]{0.48\textwidth}
        \centering
        \includegraphics[width=8.2cm]{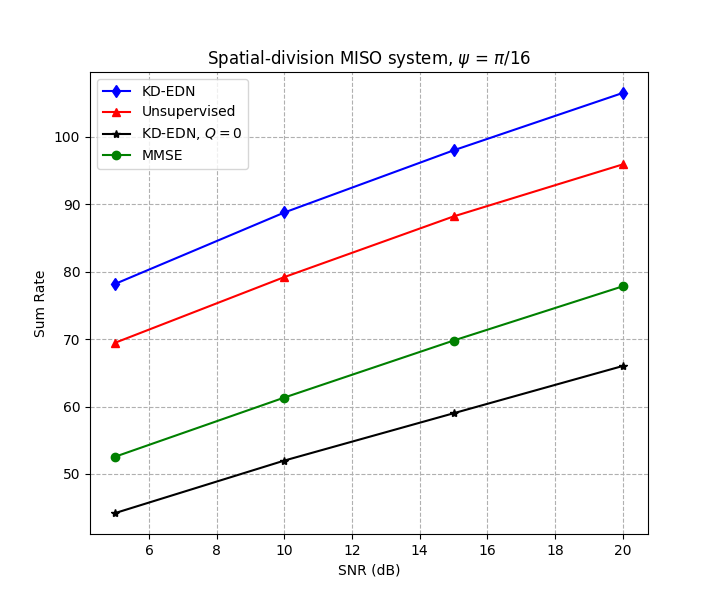}
            \caption{Spatial-division MISO system.}
	    \label{diff_snr_sd_miso_nf}
    \end{subfigure}
        \caption{Sum-rate versus SNR in near-field hybrid beamforming systems.}
        \label{diff_snr_nf}
\end{figure*}

Fig.~\ref{diff_snr_ff} and Fig.~\ref{diff_snr_nf} illustrate the sum-rate performance versus SNR for different schemes in far-field and near-field hybrid beamforming systems, respectively. The sector angle is set to $\phi=\frac{\pi}{2}$ for single-cell systems and $\psi=\frac{\pi}{16}$ for spatial-division systems. Consistent with the fully digital results in Fig.~\ref{diff_snr_miso}, the proposed \ac{KD-EDN} scheme consistently achieves the highest sum rate and outperforms all baselines in both far-field and near-field scenarios, demonstrating that the proposed \ac{KD-EDN} framework generalizes effectively across different SNR regimes and hybrid beamforming settings. 

\vspace{-3pt}
\section{conclusions} \label{sec:conclu} 

We have developed an end-to-end deep learning approach to downlink beamforming over FDD large-scale sparse \ac{MIMO} channels. A deep \ac{EDN} architecture is proposed to approximate the channel-to-latent, latent-to-beamformer, and latent-to-channel mappings, and its training is facilitated by two strategies --- semi-amortized learning and knowledge distillation. The proposed \ac{EDN} framework is extended to hybrid beamforming in both far-field and near-field channels. Extensive simulations results demonstrate consistent high-rate performances across different user deployment scenarios and different channel conditions.

\vspace{-3pt}

\appendix
We compute the gradient terms $\frac{\partial \bm{w}_Q}{\partial \bm{w}_0}$, $\frac{\partial \Bar{\bm{w}}_Q}{\partial \bm{w}_0}$, $\frac{\partial \bm{w}_Q}{\partial \Bar{\bm{w}}_0}$ and $\frac{\partial \Bar{\bm{w}}_Q}{\partial \Bar{\bm{w}}_0}$ in \eqref{theta_partial}. We begin with $\frac{\partial \bm{w}_q}{\partial \bm{w}_{q-1}}$, where $\bm{w}_q \in \mathbb{C}^{KN}$, $q\in[Q]$. Note that
\begin{align} \label{one_gradient_step}
\bm{w}_q = \bm{w}_{q-1} - \eta \nabla_{\bm{w}} L,
\end{align}
where $L$ is the loss function. We take the $a^{\text{th}}$ entry in \eqref{one_gradient_step}, and by using the Wirtinger equation $[\nabla_{\bm{w}} L]_{a} = 2 \cdot \frac{\partial L}{\partial \Bar{\bm{w}}_{q-1}(a)}$, we have
\begin{align} \label{entry_one_gradient}
\bm{w}_q(a) = \bm{w}_{q-1}(a) - 2\eta \cdot \frac{\partial L}{\partial \Bar{\bm{w}}_{q-1}(a)}.
\end{align}
According to \eqref{entry_one_gradient}, we have
\begin{align} \label{vec_one_deriv}
\frac{\partial \bm{w}_q(a)}{\partial \bm{w}_{q-1}(b)} = \delta_{a,b} - 2\eta \cdot \frac{\partial^2 L}{\partial \bm{w}_{q-1}(b) \Bar{\bm{w}}_{q-1}(a)},
\end{align}
where $a\in[KN]$ and $b\in[KN]$. Combining the entry-wise derivative in \eqref{vec_one_deriv}, we obtain the following Jacobian matrix:
\begin{align} \label{combine_one_deriv}
\frac{\partial \bm{w}_q}{\partial \bm{w}_{q-1}} = \left[\frac{\partial \bm{w}_q(a)}{\partial \bm{w}_{q-1}(b)}\right]_{a,b=1}^{KN} \in \mathbb{C}^{KN \times KN}.
\end{align}
According to the complex chain rule, we have
\begin{align} \label{induction_equ}
\left[\frac{\partial \bm{w}_q}{\partial \bm{w}_0}, \frac{\partial \Bar{\bm{w}}_q}{\partial \bm{w}_0} \right]^T =  
\underbrace{\begin{bmatrix}
\frac{\partial \bm{w}_q}{\partial \bm{w}_{q-1}} & \frac{\partial \bm{w}_q}{\partial \Bar{\bm{w}}_{q-1}} \\
\frac{\partial \Bar{\bm{w}}_q}{\partial \bm{w}_{q-1}} & \frac{\partial \Bar{\bm{w}}_q}{\partial \Bar{\bm{w}}_{q-1}} 
\end{bmatrix}}_{\mathcal{S}_q}
\cdot \left[\frac{\partial \bm{w}_{q-1}}{\partial \bm{w}_0}, \frac{\partial \Bar{\bm{w}}_{q-1}}{\partial \bm{w}_0} \right]^T.
\end{align}
Therefore we have
\begin{align} \label{induct_part_1}
\left[\frac{\partial \bm{w}_Q}{\partial \bm{w}_0}, \frac{\partial \Bar{\bm{w}}_Q}{\partial \bm{w}_0} \right]^T = \left(\Pi_{q=2}^Q \ \mathcal{S}_q \right) \cdot \left[\frac{\partial \bm{w}_1}{\partial \bm{w}_0}, \frac{\partial \Bar{\bm{w}}_1}{\partial \bm{w}_0} \right]^T.
\end{align}
Similarly, $\frac{\partial \bm{w}_Q}{\partial \Bar{\bm{w}}_0}$ and $\frac{\partial \Bar{\bm{w}}_Q}{\partial \Bar{\bm{w}}_0}$ can be computed as follows:
\begin{align} \label{induct_part_2}
\left[\frac{\partial \bm{w}_Q}{\partial \Bar{\bm{w}}_0}, \frac{\partial \Bar{\bm{w}}_Q}{\partial \Bar{\bm{w}}_0} \right]^T = \left(\Pi_{q=2}^Q \ \mathcal{S}_q \right) \cdot \left[\frac{\partial \bm{w}_1}{\partial \Bar{\bm{w}}_0}, \frac{\partial \Bar{\bm{w}}_1}{\partial \Bar{\bm{w}}_0} \right]^T.
\end{align}
All terms on the right-hand side of \eqref{induct_part_1} and \eqref{induct_part_2} can be computed using \eqref{vec_one_deriv}.

\bibliographystyle{IEEEtran}
\bibliography{IEEEabrv,bibfile}

\end{document}